\documentclass[5p,twocolumn]{elsarticle}

\usepackage{amsmath, amssymb, graphicx}
\usepackage[hidelinks]{hyperref}
\usepackage{lineno}
\usepackage{geometry}
\usepackage{booktabs}
\usepackage{xcolor}
\usepackage{bm}
\journal{J. Phys. G: Nucl. Part. Phys.}

\begin{document}

\begin{frontmatter}

\title{Form-Factor-Constrained Non-Factorized Transition-GPD Tomography of \texorpdfstring{$N\to\Delta(1232)$}{N to Delta(1232)} Multipole Structure}

\author{R. M. Marinaro III*}
\address{School of Engineering and Computing, Christopher Newport University, Newport News, VA, USA}
\address{$\mathrm{*Corresponding\:author- ralph.marinaro@cnu.edu}$}

\date{\today}

\begin{abstract}
A form-factor-constrained non-factorized transition-GPD reconstruction is developed for the \(N\to\Delta(1232)\) electromagnetic transition. The production input consists of the published CLAS magnetic multipole amplitude and electric and scalar/Coulomb quadrupole ratios. Additional external low-\(Q^2\) ratio data are retained only as a separate ratio-sector diagnostic. The electroproduction quantities are propagated replica by replica and converted to the dimensionless Jones--Scadron transition form factors \(G_M^*\), \(G_E^*\), and \(G_C^*\). AICc model selection yields a modified dipole for \(G_M^*\), a dipole for \(G_E^*\), and a cubic \(z\) expansion for \(G_C^*\). These fits define the imposed positive-\(x\) normalization \(F_\alpha(t)=2G_\alpha^*(-t)\) for factorized, correlated exponential, Regge-like, and double-distribution-inspired transition-GPD profiles. The upper boundary of the impact-parameter transform is fixed by the largest measured absolute-form-factor point, \(Q^2=6~\mathrm{GeV}^2\), so that no high-\(Q^2\) continuation beyond the measured range enters the production reconstruction. The primary transverse localization observable is therefore defined from the small-\(Q^2\) slope of the modeled transition GPD, with finite-window radial moments retained only as secondary core-shape diagnostics. For the slope-defined width, the transition form factor contributes an \(x\)-independent channel-dependent offset, while the entire \(x\) dependence is fixed by the chosen longitudinal--transverse profile ansatz. The factorized profile is \(x\)-independent and the three non-factorized families generate distinct modeled \(x\) dependences. The resulting reconstruction separates measured transition-form-factor information from model-dependent longitudinal structure and provides a controlled framework for assessing structural and covariance sensitivity in \(N\to\Delta(1232)\) transition tomography.
\end{abstract}

\begin{keyword}
Transition generalized parton distributions \sep $\Delta$(1232) resonance \sep Impact-parameter tomography \sep Transition form factors \sep Non-factorized GPDs
\end{keyword}

\end{frontmatter}

%\linenumbers

\section{Introduction}

Generalized parton distributions (GPDs) are one of the primary approaches for describing hadron structure beyond one-dimensional parton densities and elastic form factors. By connecting longitudinal momentum information with transverse spatial structure, GPDs give access to correlations that are not visible in inclusive scattering alone and have become a major theoretical framework for interpreting hard exclusive reactions, form-factor systematics, and spatial imaging of hadrons~\cite{Muller:1994ses,Ji:1997ek,Radyushkin:1996nd,Radyushkin:1997ki,Goeke:2001tz,Diehl:2003ny,Belitsky:2005qn}. In the zero-skewness limit, the transverse Fourier interpretation of GPDs gives a particularly intuitive representation in impact-parameter space, where form-factor information can be connected to spatial localization of partonic strength~\cite{Burkardt:2000za,Burkardt:2002hr}. This spatial interpretation has been widely used for elastic nucleon structure, but its extension to transition systems remains less developed despite its potential to illuminate how the internal structure of a baryon is reorganized during excitation.

Transition GPDs generalize the elastic case by describing matrix elements between different hadronic states. They are therefore naturally suited to resonance physics, where the relevant question is not only how charge, magnetization, or current distributions are arranged within a single baryon, but how those distributions are rearranged when a nucleon is excited into a higher-lying state. The \(N\to\Delta(1232)\) transition is an especially important benchmark for this focus. The \(\Delta(1232)\) is the lowest-lying nucleon resonance, dominates pion electroproduction in the first resonance region, and has long served as a testing ground for models of baryon deformation, meson-cloud effects, and the interplay between quark-core and long-range pion dynamics~\cite{Jones:1972ky,Drechsel:2007if,Aznauryan:2011qj}. Its electromagnetic excitation is dominated by the magnetic dipole transition, while the smaller electric and scalar/Coulomb quadrupole components carry information about deformation and longitudinal response. The coexistence of a dominant transition strength with subleading deformation-sensitive amplitudes makes the \(N\to\Delta(1232)\) channel a particularly useful system for developing transition tomography.

This work is also situated within the broader transition-GPD program for baryon resonances. Recent reviews and white-paper studies have emphasized transition GPDs as a developing framework for baryon-resonance tomography, connecting exclusive electroproduction observables with spatial, spin, and mechanical structure in non-diagonal hadronic matrix elements~\cite{Diehl:2025TransitionGPDWhitePaper}. For the specific \(N\to\Delta\) case, recent formal work has clarified that a complete leading-twist transition-GPD definition requires a full set of independent covariant structures, including structures not contained in earlier form-factor-inspired parametrizations~\cite{Kim:2025CompleteDefinition}. Related studies have developed the light-front and multipole interpretation of \(N\to\Delta\) transition GPDs, including monopole, dipole, and quadrupole components in the transverse plane~\cite{Kim:2026MultipoleStructure}, transition angular-momentum densities~\cite{Kim:2023QCDAM}, and tensor-polarized transition parton densities~\cite{Kim:2025TensorPolarized}. The present work should therefore be understood as a phenomenological, form-factor-constrained reconstruction complementary to this formal program, rather than as a complete operator-level parametrization of all \(N\to\Delta\) transition GPDs. The present analysis is restricted to the electromagnetic vector-current sector and the associated \(H_M\), \(H_E\), and \(H_C\) transition structures. Axial-vector, tensor, and energy--momentum-tensor transition GPDs are outside the scope of the present reconstruction.

A substantial empirical and theoretical literature has established the importance of the \(N\to\Delta(1232)\) transition form factors. Pion electroproduction measurements have provided detailed constraints on the magnetic, electric, and scalar/Coulomb multipole content over a broad range of momentum transfer~\cite{Frolov:1998pw,Joo:2001tw,Ungaro:2006df,Aznauryan:2009mx}. These measurements have been interpreted through phenomenological analyses, dynamical reaction models, constituent-quark calculations, chiral effective approaches, and lattice-QCD studies, each emphasizing different aspects of the transition current~\cite{Sato:2000jf,Kamalov:2001qg,Pascalutsa:2006up,Alexandrou:2007dt,Gail:2005gz,Ramalho:2016buz}. A recurring conclusion is that the small quadrupole amplitudes and the low-\(Q^2\) behavior are sensitive to physics beyond a simple compact quark core, including pion-cloud contributions and constraints associated with current conservation and Siegert behavior~\cite{Pascalutsa:2005ts,Vanderhaeghen:2007zz,Ramalho:2016buz}. These observations motivate a framework that can retain empirical form-factor constraints while allowing the spatial structure of the transition to be interrogated in a controlled and reproducible way.

The precursor to the present work introduced such a baseline by reconstructing a transition-GPD model for the \(\Delta(1232)\) from a fitted helicity-amplitude input and by using impact-parameter diagnostics to connect amplitude behavior with transverse spatial localization~\cite{Marinaro:2025TransitionGPD}. That study was intentionally minimal. It used a factorized longitudinal--transverse ansatz, imposed the relevant form-factor normalization, and demonstrated that a transparent amplitude-to-space pipeline could be implemented with controlled uncertainty propagation. Its main value was to establish a reproducible bridge between transition-amplitude data and spatial diagnostics. At the same time, the factorized construction also exposed an important limitation. If the longitudinal and transverse dependences are separated by assumption, then the resulting transverse profile cannot contain genuine \(x\)-dependent localization beyond the imposed longitudinal weighting. A more realistic transition-tomography framework must therefore allow controlled correlations between momentum fraction and momentum transfer while maintaining consistency with the measured transition form factors.

The present paper develops that next step. Rather than treating the transition through a single helicity-amplitude input, it begins from the directly published \(\Delta(1232)\) magnetic multipole and quadrupole-ratio data from CLAS. The raw electroproduction quantities are propagated replica by replica and converted to the dimensionless Jones--Scadron transition form factors \(G_M^*\), \(G_E^*\), and \(G_C^*\) before the transition-GPD construction is introduced. The low-\(Q^2\) external measurements remain confined to a separate ratio-sector diagnostic and do not create additional absolute transition-form-factor points. The resulting form factors provide the empirical momentum-transfer information used to normalize the phenomenological transition-GPD profiles, while the unresolved longitudinal dependence is represented by controlled factorized and non-factorized profile families. This separation keeps the data-constrained electromagnetic transition strength distinct from the modeled \(x\)--\(t\) structure and allows profile-dependent transverse localization to be studied under a common empirical normalization.

This extension is motivated by both phenomenology and interpretation. From the phenomenological side, the available electroproduction data constrain the momentum-transfer dependence of transition strength much more directly than they constrain the full \(x\)-dependence of a transition GPD. Any spatial reconstruction must therefore distinguish between data-driven form-factor information and model-controlled longitudinal structure. From the interpretive side, the comparison of factorized, correlated, Regge-like, and double-distribution-inspired profile families provides a controlled way to ask which spatial features are robust consequences of the form-factor input and which are structural consequences of the assumed \(x\)--\(t\) correlation. This distinction is essential if transition tomography is to become useful for comparing empirical extractions, lattice-QCD calculations, and hadronic models on common spatial footing.

The paper is organized as follows. Section~\ref{sec:data} summarizes the published \(\Delta(1232)\) inputs, the construction of the derived raw multipoles, and their conversion to Jones--Scadron transition form factors. Section~\ref{sec:fits} discusses the transition-form-factor fits and model-selection diagnostics. Section~\ref{sec:gpd} introduces the form-factor-constrained transition-GPD construction and the profile families used to test factorization breaking. Section~\ref{sec:tomography} presents the band-limited impact-parameter representation, while Section~\ref{sec:localization} introduces the slope-defined transverse width together with finite-window mean-radius and shape diagnostics. Section~\ref{sec:multipole_kernels} gives the multipole-resolved radial kernels and clarifies their relation to formal transition-density constructions. Section~\ref{sec:discussion} discusses the uncertainty decomposition, higher transverse-shape moments, profile dependence, and physical interpretation. Section~\ref{sec:summary} summarizes the main conclusions and outlook.

\section{Published \texorpdfstring{$\Delta(1232)$}{Delta(1232)} inputs, derived multipoles, and Jones--Scadron transition form factors}
\label{sec:data}

The empirical starting point of the present analysis is the set of directly published \(\Delta(1232)\) electro-excitation observables from the CLAS single-pion electroproduction analysis of Aznauryan \emph{et al.}~\cite{Aznauryan:2009mx}. These observables are explicitly tabulated for the \(\Delta(1232)\) channel. For this resonance, the CLAS analysis reports the magnetic multipole amplitude \(\mathrm{Im}\,M_{1+}^{3/2}\) at \(W=1.232~\mathrm{GeV}\), together with the electric and scalar/Coulomb quadrupole ratios \(R_{EM}\) and \(R_{SM}\). The primary CLAS inputs are organized into three data sets. The first is the magnetic transition strength \(\mathrm{Im}\,M_{1+}^{3/2}\), reported in units of \(\sqrt{\mu{\rm b}}\). The second and third are the quadrupole ratios \(R_{EM}\) and \(R_{SM}\), reported in percent. The ratio observables are defined by
\begin{align}
  R_{EM}(Q^2)=100\,\frac{\mathrm{Im}\,E_{1+}^{3/2}(Q^2)}
  {\mathrm{Im}\,M_{1+}^{3/2}(Q^2)},
  \\
  R_{SM}(Q^2)=100\,\frac{\mathrm{Im}\,S_{1+}^{3/2}(Q^2)}
  {\mathrm{Im}\,M_{1+}^{3/2}(Q^2)} .
  \label{eq:ratios}
\end{align}
The notation follows the conventional pion-electroproduction multipole language for the \(P_{33}(1232)\) resonance, in which the magnetic dipole transition gives the dominant contribution and the electric and scalar/Coulomb quadrupole ratios quantify deformation-sensitive admixtures~\cite{Jones:1972ky,Drechsel:2007if,Pascalutsa:2006up}. In this input-level discussion, the shorthand labels \(M\), \(E\), and \(S\) refer only to the magnetic, electric, and scalar/Coulomb electroproduction multipoles. After conversion to electromagnetic transition form factors, the corresponding Jones--Scadron sectors are denoted \(M\), \(E\), and \(C\), corresponding to \(G_M^*\), \(G_E^*\), and \(G_C^*\), respectively.

\begin{figure*}[t]
    \centering
    \includegraphics[width=\textwidth]{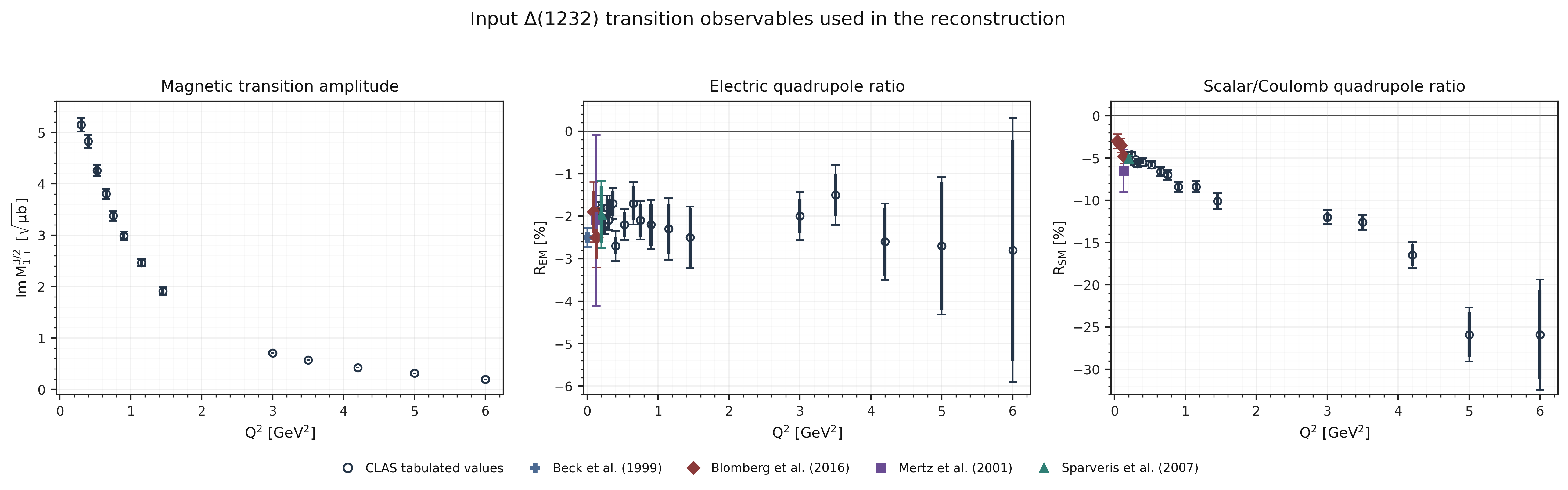}
    \caption{Published \(\Delta(1232)\) input observables used in the reconstruction. The magnetic panel shows the directly tabulated CLAS \(\mathrm{Im}\,M_{1+}^{3/2}\) values, while the \(R_{EM}\) and \(R_{SM}\) panels show the CLAS ratio data together with explicitly labeled low-\(Q^2\) ratio-sector references.}
    \label{fig:clas_input_data}
\end{figure*}

This direct use of the CLAS tables avoids an ambiguity that is important for a transition-tomography analysis. The analysis begins from the tabulated \(\mathrm{Im}\,M_{1+}^{3/2}\), \(R_{EM}\), and \(R_{SM}\) values and derives the corresponding electric and scalar/Coulomb multipole amplitudes only where the magnetic and ratio inputs share the same \(Q^2\) point. No interpolation is used to create derived multipole amplitudes on the magnetic grid. This preserves a transparent chain from the published observables to the raw derived multipoles, their Jones--Scadron conversion, and the transition form factors used in the subsequent fits and spatial reconstruction. Figure~\ref{fig:clas_input_data} displays the input observables used to anchor the reconstruction. The magnetic panel contains only same-convention magnetic-amplitude points, while the quadrupole-ratio panels include the published CLAS ratios and the external low-\(Q^2\) ratio-sector references discussed below. This separation is essential because the ratio measurements and the magnetic multipole amplitudes are not interchangeable without a documented convention and unit conversion.

\begin{table*}[t]
\centering
\caption{Primary published inputs used in the \(\Delta(1232)\) multipole reconstruction. The magnetic amplitude is used as the normalization channel for the derived electric and scalar/Coulomb multipoles. The ratio points enter the derivation only at \(Q^2\) values that coincide with the magnetic grid, while additional low-\(Q^2\) ratio references are used only in the ratio-sector validation.}
\label{tab:published_inputs}
\begin{tabular}{llll}
\toprule
Input observable & Source & Units & Role in this analysis \\
\midrule
\(\mathrm{Im}\,M_{1+}^{3/2}\) & CLAS Table VI~\cite{Aznauryan:2009mx} & \(\sqrt{\mu{\rm b}}\) & Magnetic transition amplitude \\
\(R_{EM}\) & CLAS Table VII~\cite{Aznauryan:2009mx} & percent & Electric-to-magnetic quadrupole ratio \\
\(R_{SM}\) & CLAS Table VIII~\cite{Aznauryan:2009mx} & percent & Scalar/Coulomb-to-magnetic quadrupole ratio \\
Low-\(Q^2\) \(R_{EM}\), \(R_{SM}\) & Refs.~\cite{Beck:2000, Mertz:2001, Sparveris:2007, Blomberg:2016} & percent & Ratio-sector constraints only \\
\bottomrule
\end{tabular}
\end{table*}

\subsection{Low-\texorpdfstring{$Q^2$}{Q2} ratio-sector constraints}

The low-\(Q^2\) behavior of the \(N\to\Delta(1232)\) quadrupole ratios is phenomenologically important because this is the region where pion-cloud dynamics, current-conservation constraints, and Siegert behavior are expected to have the greatest impact~\cite{Pascalutsa:2005ts,Pascalutsa:2006up,Ramalho:2016buz}. To retain sensitivity to this physics without introducing inconsistent same-unit multipole points, additional low-\(Q^2\) information is incorporated only in the \(R_{EM}\) and \(R_{SM}\) ratio sector. These external references include the Mainz photoproduction extraction of the real-photon \(E2/M1\) ratio~\cite{Beck:2000}, the Bates/OOPS measurement at \(Q^2=0.126~\mathrm{GeV}^2\)~\cite{Mertz:2001}, the MAMI measurement at \(Q^2=0.20~\mathrm{GeV}^2\)~\cite{Sparveris:2007}, and the Hall A low-momentum-transfer measurements extending the Coulomb quadrupole information to very low \(Q^2\)~\cite{Blomberg:2016}.

The role of these points is intentionally limited. They are used to show how the quadrupole-ratio sector behaves when the low-\(Q^2\) region is included, and they provide an empirical check on the ratio-level trends relevant to pion-cloud and Siegert-region interpretations. They are not used to generate additional \(\mathrm{Im}\,M_{1+}^{3/2}\), \(\mathrm{Im}\,E_{1+}^{3/2}\), or \(\mathrm{Im}\,S_{1+}^{3/2}\) multipole amplitudes in the same units as the CLAS magnetic table. In particular, real-part ratio extractions and magnetic amplitudes quoted in different conventions are not converted into the \(\sqrt{\mu{\rm b}}\) CLAS convention unless a complete and explicit conversion is supplied. This convention choice keeps the data provenance auditable and prevents the form-factor fits in Section~\ref{sec:fits} from being constrained by mixed-unit inputs.

\subsection{Monte Carlo propagation into the raw multipole amplitudes}

The derived electric and scalar/Coulomb multipole amplitudes are obtained by propagating the measured magnetic amplitude and the two ratio observables replica by replica. For each published table, the statistical and model uncertainties are retained as separate inputs. Because the published tables do not provide a complete point-to-point covariance matrix for the quoted model component, the central analysis combines the statistical and model uncertainties in quadrature point by point. A second limiting prescription treats the quoted model uncertainty within each table as fully correlated across momentum-transfer points. The latter is used as an explicit covariance-sensitivity test rather than as an alternative central result. The quantitative effect of this choice on the form-factor fits and spatial observables is reported in Section~\ref{sec:discussion}.

At each \(Q^2\) point on the magnetic grid for which the corresponding ratio value is directly tabulated, the derived multipoles are constructed as
\begin{align}
  E_{1+}^{3/2}(Q^2)=
  \frac{R_{EM}(Q^2)}{100}\,M_{1+}^{3/2}(Q^2),\\
  S_{1+}^{3/2}(Q^2)=
  \frac{R_{SM}(Q^2)}{100}\,M_{1+}^{3/2}(Q^2).
  \label{eq:derived_ES}
\end{align}
The multiplication is performed for each Monte Carlo replica rather than only for the central values. This treatment preserves the common magnetic normalization shared by the \(E\) and \(S\) amplitudes and therefore retains the induced correlations among the derived multipoles. The resulting samples define the raw multipole central values, uncertainties, and covariance structure. These quantities are retained for experimental provenance and for the diagnostic amplitude combinations below, but they are not used directly as the transition-form-factor normalization of the GPD construction.

\begin{figure*}[h!]
    \centering
    \includegraphics[width=\textwidth]{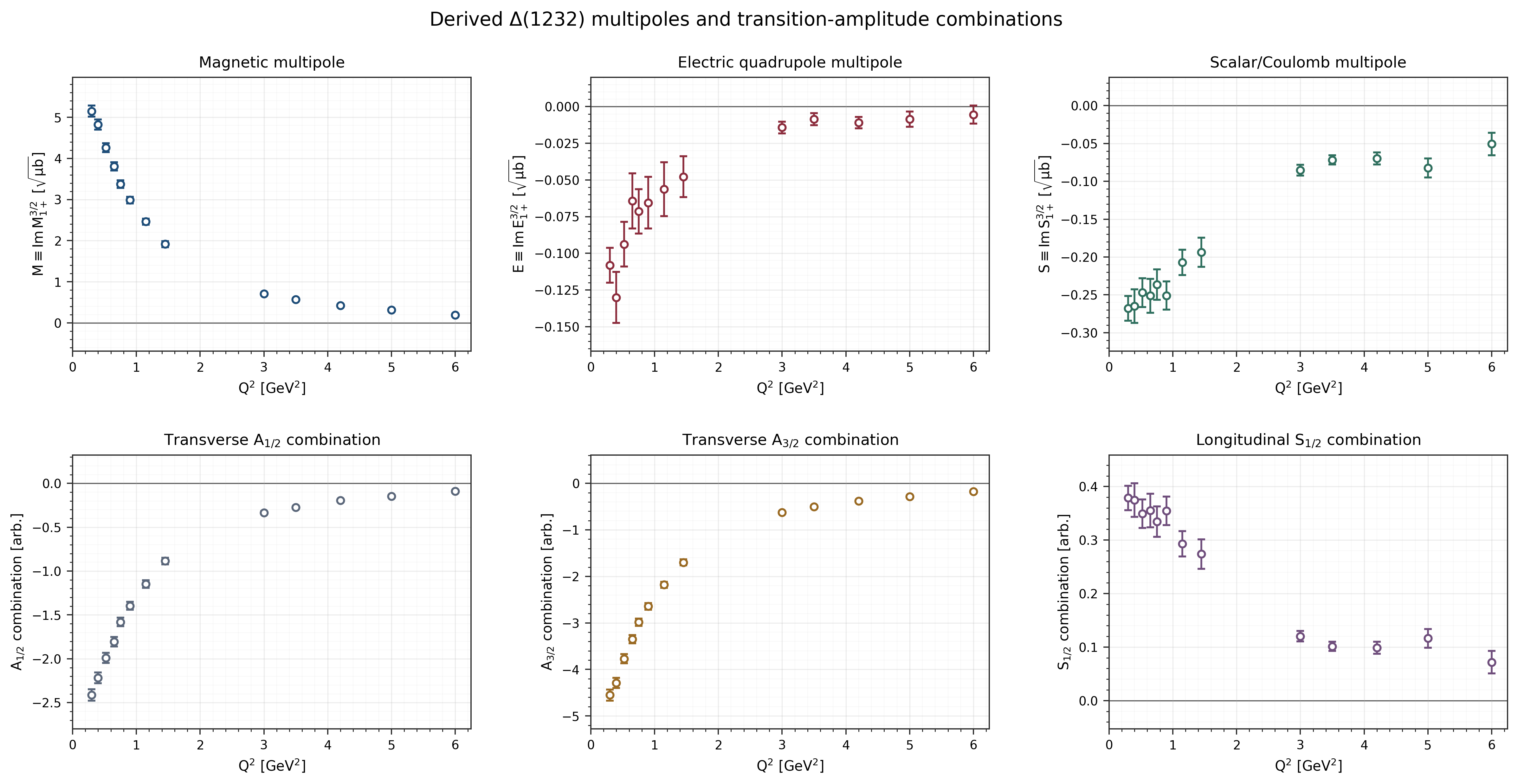}
    \caption{Derived \(\Delta(1232)\) electroproduction multipoles and diagnostic transition-amplitude combinations obtained from the directly tabulated CLAS magnetic amplitude and quadrupole ratios. The upper row shows the derived \(M\), \(E\), and \(S\) multipoles in \(\sqrt{\mu{\rm b}}\). The lower row shows the \(A_{1/2}\), \(A_{3/2}\), and \(S_{1/2}\)-like diagnostic linear combinations in arbitrary units. No absolute helicity-amplitude normalization is assigned to these diagnostic panels. The electric and scalar/Coulomb amplitudes are propagated replica by replica so that the common magnetic normalization is retained in the covariance structure.}
    \label{fig:derived_multipoles}
\end{figure*}

For comparison with the familiar helicity-amplitude structure, we form three linear diagnostic combinations from the derived electroproduction multipoles,
\begin{align}
  A_{1/2}^{\mathrm{comb}} &\propto -\frac{1}{2}(M+3E),\\
  A_{3/2}^{\mathrm{comb}} &\propto -\frac{\sqrt{3}}{2}(M-E),\\
  S_{1/2}^{\mathrm{comb}} &\propto -\sqrt{2}\,S .
  \label{eq:amplitude_combinations}
\end{align}
The proportionality signs are intentional. These quantities are not presented as absolutely normalized physical helicity amplitudes, because the additional kinematic and convention-dependent normalization required for such an identification is not imposed here. They are retained only as arbitrary-unit diagnostics of the relative linear combinations of the measured \(M\), \(E\), and \(S\) multipoles. Accordingly, the three lower panels of Figure~\ref{fig:derived_multipoles} are labeled in arbitrary units. These diagnostic combinations do not enter the Jones--Scadron transition-form-factor fits, the transition-GPD normalization, or the impact-parameter tomography. The physical electromagnetic quantities used in the subsequent analysis are instead the dimensionless Jones--Scadron form factors constructed below.

\subsection{Conversion to Jones--Scadron transition form factors}

The transition-GPD first-moment relations are conventionally expressed in terms of electromagnetic \(N\to\Delta\) transition form factors rather than directly in terms of pion-electroproduction multipoles. We therefore convert the published resonance multipoles to the dimensionless Jones--Scadron form factors \(G_M^*\), \(G_E^*\), and \(G_C^*\) before carrying out the form-factor fits or the transverse reconstruction.

For the magnetic sector, the resonance multipole is first converted to the Ash magnetic transition form factor using the relation given by Aznauryan \emph{et al.}~\cite{Aznauryan:2009mx},
\begin{equation}
G_{M,\mathrm{Ash}}^*(Q^2)
=
\frac{M_N}{k_\Delta(Q^2)}
\left[
\frac{8q_\pi\Gamma_\Delta}
{3\alpha_{\rm em}}
\right]^{1/2}
\mathrm{Im}\,M_{1+}^{3/2}(Q^2),
\label{eq:GM_Ash_conversion}
\end{equation}
where the tabulated multipole amplitude is first converted from \(\sqrt{\mu{\rm b}}\) to natural units using
\(1~\mathrm{GeV}^{-2}=389.3793721~\mu{\rm b}\).
The resonance-frame momenta are
\begin{align}
k_\Delta(Q^2)
&=
\frac{Q_+(Q^2)Q_-(Q^2)}
{2M_\Delta},
\\
Q_\pm(Q^2)
&=
\sqrt{(M_\Delta\pm M_N)^2+Q^2},
\label{eq:delta_kinematics}
\end{align}
and
\begin{equation}
q_\pi=
\frac{
\sqrt{
\left[M_\Delta^2-(M_N+m_\pi)^2\right]
\left[M_\Delta^2-(M_N-m_\pi)^2\right]
}
}
{2M_\Delta}.
\label{eq:qpi_delta}
\end{equation}
The numerical implementation uses \(M_N=0.9382721~\mathrm{GeV}\), \(M_\Delta=1.232~\mathrm{GeV}\), \(\Gamma_\Delta=0.118~\mathrm{GeV}\), \(m_\pi=0.1349768~\mathrm{GeV}\), and \(\alpha_{\rm em}^{-1}=137.035999084\).

The Ash form factor is then converted to the Jones--Scadron convention~\cite{Aznauryan:2009mx,Jones:1972ky},
\begin{equation}
G_M^*(Q^2)
=
G_{M,\mathrm{Ash}}^*(Q^2)
\sqrt{
1+\frac{Q^2}
{(M_\Delta+M_N)^2}
}.
\label{eq:Ash_JS_conversion}
\end{equation}
At the resonance position the quadrupole ratios are related to the Jones--Scadron form factors by~\cite{Jones:1972ky,Pascalutsa:2006up}
\begin{align}
R_{EM}(Q^2)
&=
-\frac{G_E^*(Q^2)}
{G_M^*(Q^2)},
\\
R_{SM}(Q^2)
&=
-\frac{Q_+(Q^2)Q_-(Q^2)}
{4M_\Delta^2}
\frac{G_C^*(Q^2)}
{G_M^*(Q^2)}.
\label{eq:JS_ratio_relations}
\end{align}
Because the experimental ratios are tabulated in percent, the electric and Coulomb form factors used in the numerical analysis are therefore
\begin{align}
G_E^*(Q^2)
&=
-\frac{R_{EM}(Q^2)}{100}\,
G_M^*(Q^2),
\\
G_C^*(Q^2)
&=
-\frac{4M_\Delta^2}
{Q_+(Q^2)Q_-(Q^2)}
\frac{R_{SM}(Q^2)}{100}\,
G_M^*(Q^2).
\label{eq:GE_GC_conversion}
\end{align}
All conversions are performed replica by replica. Consequently, the common magnetic normalization and the correlations induced by the measured quadrupole ratios propagate directly into the \(G_E^*\) and \(G_C^*\) covariance matrices. As a convention audit, the converted magnetic form factor was compared with the independently published high-\(Q^2\) quantity \(100G_M^*/(3G_D)\) of Ref.~\cite{Ungaro:2006df}. At the five overlapping points \(Q^2=3.0,\ 3.5,\ 4.2,\ 5.0,\) and \(6.0~\mathrm{GeV}^2\), the present conversion gives \(62.35,\ 59.83,\ 55.36,\ 51.54,\) and \(40.11\%\), respectively, differing from the published values by no more than \(1.57\) percentage points. This provides an independent check of the normalization and convention used in the conversion.

The quantities fitted in Section~\ref{sec:fits} and used throughout the transition-GPD construction are therefore the dimensionless Jones--Scadron form factors \(G_M^*\), \(G_E^*\), and \(G_C^*\). The original \(\sqrt{\mu{\rm b}}\) multipoles remain part of the documented input layer but are not themselves identified with the transition form factors entering the GPD normalization.

\section{\texorpdfstring{$Q^2$}{Q2}-dependent transition form factors and model selection}
\label{sec:fits}

The Jones--Scadron transition form factors \(G_M^*\), \(G_E^*\), and \(G_C^*\) constructed in Section~\ref{sec:data} provide the empirical electromagnetic input for the transition-GPD reconstruction. Before introducing any longitudinal profile model, their measured momentum-transfer dependence is represented by a compact library of candidate functions and compared using finite-sample information criteria. This step separates the data-constrained \(Q^2\) dependence of the physical transition form factors from the model-controlled \(x\)-dependent structure introduced in Section~\ref{sec:gpd}.

Throughout this section the channel label \(\alpha\) denotes
\[
\alpha\in\{M,E,C\},
\]
corresponding to the magnetic dipole, electric quadrupole, and Coulomb quadrupole Jones--Scadron form factors \(G_M^*\), \(G_E^*\), and \(G_C^*\), respectively. These fitted quantities are dimensionless. The channel-by-channel fits use the covariance matrices produced by the replica-by-replica conversion described in Section~\ref{sec:data}, thereby retaining the within-channel correlations inherited from the common magnetic normalization and the measured quadrupole ratios. The central analysis combines the published statistical and model uncertainties in quadrature point by point because a complete point-to-point covariance matrix for the model component is not available. A fully correlated within-table model-error prescription is evaluated separately as a limiting covariance cross-check. The central \(M\), \(E\), and \(C\) fits are performed independently and therefore do not use the off-diagonal covariance blocks between different transition sectors, although those cross-channel correlations are retained by the common replica ensemble. A joint \(M/E/C\) fit using the full replica-derived covariance block is also performed as a dedicated sensitivity test and is discussed in Section~\ref{sec:discussion}. The data-level propagation uses \(N_{\rm rep}=5000\) replicas and a fixed random seed.

\subsection{Fit-model library}

The fit library is intentionally compact. The goal is not to introduce a global phenomenological model of the \(N\to\Delta\) transition current, but to obtain smooth, auditable representations of the measured Jones--Scadron form-factor dependence that can be propagated into impact-parameter space. The simplest baseline is a dipole form,
\begin{equation}
  G_{\alpha,\mathrm{dip}}^*(Q^2)=
  \frac{A_{\alpha,0}}{(1+Q^2/\Lambda_\alpha^2)^2},
  \qquad
  \alpha\in\{M,E,C\} .
  \label{eq:dipole}
\end{equation}
Dipole-like behavior is a standard phenomenological reference in form-factor studies and provides a minimal parametrization of a localized transition current~\cite{Sachs:1962zzc,Kelly:2004hm}. In the present work it also provides a direct point of comparison with the precursor analysis, where a dipole representation supplied the first amplitude-to-space bridge. To allow modest deviations from a pure dipole falloff, a modified dipole form is included,
\begin{equation}
  G_{\alpha,\mathrm{mod}}^*(Q^2)=
  \frac{A_{\alpha,0}(1+c_\alpha Q^2)}
  {(1+Q^2/\Lambda_\alpha^2)^2} .
  \label{eq:modified_dipole}
\end{equation}
The additional linear term lets the data adjust the curvature of the form factor without introducing a high-order polynomial that can become unstable outside the measured region. This model is therefore useful for testing whether the inferred spatial diagnostics are sensitive to small deviations from the simplest dipole shape. A model-independent comparison class is provided by truncated \(z\)-expansion forms~\cite{Hill:2010yb,Bhattacharya:2011ah},
\begin{equation}
  G_{\alpha,z}^*(Q^2)=\sum_{k=0}^K a_{\alpha,k} z(Q^2)^k ,
  \label{eq:zexpansion}
\end{equation}

where the conformal variable is defined by
\begin{equation}
z(Q^2;t_{\rm cut},t_0)=
\frac{
\sqrt{t_{\rm cut}+Q^2}
-
\sqrt{t_{\rm cut}-t_0}
}{
\sqrt{t_{\rm cut}+Q^2}
+
\sqrt{t_{\rm cut}-t_0}
}.
\label{eq:zvariable}
\end{equation}
For the electromagnetic \(N\to\Delta\) vector transition we use
\begin{equation}
t_{\rm cut}=4m_{\pi^\pm}^2
=0.07792~\mathrm{GeV}^2,
\qquad
t_0=0 .
\end{equation}
The value \(t_{\rm cut}=4m_{\pi^\pm}^2\) corresponds to the first normal two-pion isovector electromagnetic \(t\)-channel cut. The choice \(t_0=0\) gives \(z(Q^2=0)=0\), so that the constant coefficient of the truncated expansion is the photon-point value. Dispersive treatments of the \(N\leftrightarrow\Delta\) transition also contain anomalous analytic structure associated with the unstable \(\Delta\)~\cite{Aung:2025DeltaTFF}. The present finite-order \(z\) expansion is therefore used only as an empirical parametrization of the spacelike data-supported region and is not intended as a complete dispersive representation. The cubic and quartic versions are retained as controlled comparison models rather than as high-order extrapolation tools.

The absolute transition-form-factor parametrizations are treated as empirical representations of the data-supported spacelike region and are not assigned asymptotic validity at arbitrarily large \(Q^2\). In particular, the production impact-parameter reconstruction does not extrapolate the selected \(G_M^*\), \(G_E^*\), or \(G_C^*\) fits above the largest absolute-form-factor input point, \(Q^2=6~\mathrm{GeV}^2\). Continuations beyond that boundary are evaluated only as cutoff-sensitivity diagnostics. This restriction is especially important for a finite-order \(z\) expansion, whose bounded conformal variable does not by itself enforce the large-\(Q^2\) counting behavior required of a physical transition form factor. A low-\(Q^2\) Siegert/pion-cloud-inspired parametrization is retained separately for the \(R_{EM}\) and \(R_{SM}\) ratio-sector diagnostic. The auxiliary neutron electric-form-factor shape entering this diagnostic uses the Galster parametrization~\cite{Galster:1971kv}. Its physical motivation is strongest in the very-low-\(Q^2\) quadrupole ratios, where long-range pion dynamics and current-conservation constraints are expected to be most visible~\cite{Pascalutsa:2005ts,Pascalutsa:2006up,Ramalho:2016buz}. It is therefore not included in the candidate library used for the absolute Jones--Scadron \(G_M^*\), \(G_E^*\), and \(G_C^*\) fits. The absolute-form-factor library consists of the dipole, modified-dipole, cubic \(z\)-expansion, and quartic \(z\)-expansion models. This separation prevents a low-\(Q^2\) ratio-sector sensitivity ansatz from being reinterpreted as a general absolute transition-form-factor parametrization.

\subsection{Correlated residuals and information criteria}

For each channel and model, the best-fit parameters are determined by minimizing a covariance-weighted residual. If \(\bm{y}_\alpha\) denotes the vector of derived Jones--Scadron form-factor values in channel \(\alpha\), \(\bm{f}_\alpha\) denotes the corresponding model prediction, and \(C_\alpha\) is the covariance matrix obtained from the replica propagation, the objective function is
\begin{equation}
  \chi^2_\alpha =
  \left(\bm{y}_\alpha-\bm{f}_\alpha\right)^T
  C_\alpha^{-1}
  \left(\bm{y}_\alpha-\bm{f}_\alpha\right).
  \label{eq:chi2}
\end{equation}
The covariance matrix is symmetrized and regularized when necessary to ensure numerical stability. This treatment preserves the uncertainty scale of the derived transition form factors while avoiding the assumption that all points are independent after the common magnetic normalization has been propagated into \(G_E^*\) and \(G_C^*\). The resulting fits are compared using the Akaike information criterion, the finite-sample corrected Akaike criterion, and the Bayesian information criterion,
\begin{align}
  \mathrm{AIC}= \chi^2+2k,\\\nonumber\\
  \mathrm{AICc}=\mathrm{AIC}
  +\frac{2k(k+1)}{n-k-1},\\\nonumber\\
  \mathrm{BIC}= \chi^2+k\ln n ,
  \label{eq:information_criteria}
\end{align}
where \(k\) is the number of fit parameters and \(n\) is the number of fitted data points~\cite{Akaike:1974,Schwarz:1978,Hurvich:1989,Burnham:2002}. For each of the absolute \(M\), \(E\), and \(C\) transition-form-factor fits, \(n=13\), corresponding to the 13 matched \(Q^2\) points on the CLAS magnetic grid. Table~\ref{tab:model_selection} lists \(n=13\) explicitly for every entry. The AICc value is used as the primary selection criterion because the number of fitted points in each transition-form-factor channel is modest. Akaike weights are computed within each channel to quantify relative support among the candidate models. The model-selection diagnostics are reported in Table~\ref{tab:model_selection}, including all absolute-form-factor candidate models for \(M\), \(E\), and \(C\).

\begin{table*}[t]
\centering
\caption{Model-selection diagnostics for the Jones--Scadron \(G_M^*\), \(G_E^*\), and \(G_C^*\) transition-form-factor fits. The selected model in each channel is the minimum-AICc entry within the absolute-form-factor candidate family. \(\Delta\)AICc and \(\Delta\)BIC are evaluated independently within each channel. The number of fitted points is \(n=13\) for every entry. The \(z\)-expansion candidates use \(t_{\rm cut}=4m_{\pi^\pm}^2\) and \(t_0=0\).}
\label{tab:model_selection}
\begin{tabular}{llllrrrrrrl}
\toprule
Channel & Model & \(k\) & \(n\) & \(\chi^2/\nu\) & AICc & \(\Delta\)AICc & BIC & \(\Delta\)BIC & Weight & Selected \\
\midrule
\(M\) & modified dipole & 3 & 13 & 0.43 & 12.94 & 0.00 & 11.97 & 0.00 & 0.968 & yes \\
\(M\) & quartic \(z\) expansion & 5 & 13 & 0.26 & 20.62 & 7.68 & 14.88 & 2.91 & 0.021 & -- \\
\(M\) & cubic \(z\) expansion & 4 & 13 & 0.99 & 21.93 & 8.99 & 19.19 & 7.22 & 0.011 & -- \\
\(M\) & dipole & 2 & 13 & 6.76 & 79.61 & 66.67 & 79.54 & 67.57 & \(<0.001\) & -- \\
\midrule
\(E\) & dipole & 2 & 13 & 0.55 & 11.24 & 0.00 & 11.17 & 0.00 & 0.763 & yes \\
\(E\) & modified dipole & 3 & 13 & 0.52 & 13.85 & 2.61 & 12.88 & 1.71 & 0.207 & -- \\
\(E\) & cubic \(z\) expansion & 4 & 13 & 0.53 & 17.81 & 6.57 & 15.07 & 3.90 & 0.029 & -- \\
\(E\) & quartic \(z\) expansion & 5 & 13 & 0.56 & 23.04 & 11.80 & 17.29 & 6.12 & 0.002 & -- \\
\midrule
\(C\) & cubic \(z\) expansion & 4 & 13 & 1.18 & 23.62 & 0.00 & 20.88 & 0.01 & 0.727 & yes \\
\(C\) & quartic \(z\) expansion & 5 & 13 & 1.01 & 26.62 & 3.00 & 20.87 & 0.00 & 0.163 & -- \\
\(C\) & modified dipole & 3 & 13 & 1.87 & 27.40 & 3.78 & 26.43 & 5.56 & 0.110 & -- \\
\(C\) & dipole & 2 & 13 & 3.00 & 38.18 & 14.56 & 38.11 & 17.24 & 0.001 & -- \\
\bottomrule
\end{tabular}
\end{table*}

For uncertainty propagation, the covariance matrix of the selected fit parameters is used to generate parameter replicas. These replicas are evaluated over a continuous \(Q^2\) grid to form the median fit and the central \(68\%\) uncertainty band. The same sampling is later propagated through the GPD and impact-parameter calculations. In this way, the tomography figures retain the uncertainty associated with the measured transition form factors rather than only displaying central fit curves.

\begin{figure*}[h!]
    \centering
    \includegraphics[width=\textwidth]{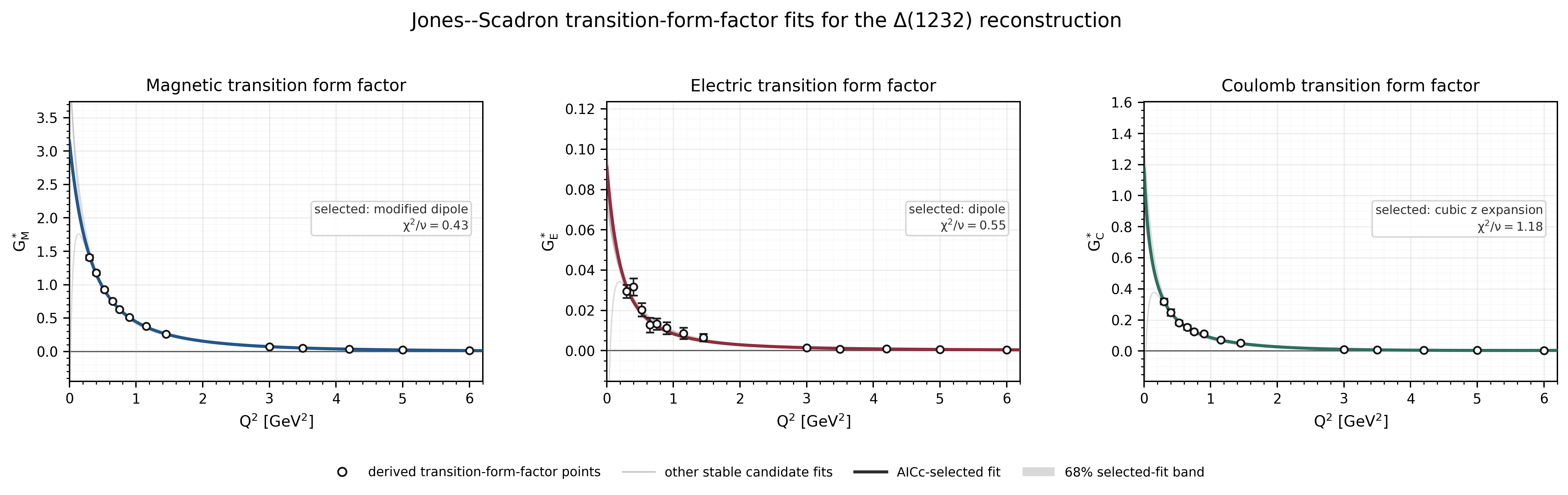}
    \caption{Jones--Scadron transition-form-factor fits used in the \(\Delta(1232)\) reconstruction. Points show derived dimensionless \(G_M^*\), \(G_E^*\), and \(G_C^*\) values with replica-propagated uncertainties. The dark curve in each panel shows the AICc-selected fit, and the shaded region gives the propagated \(68\%\) uncertainty band of the selected fit.}
    \label{fig:form_factor_model_comparison}
\end{figure*}

\subsection{Transition-form-factor results}

The model comparison selects different functional forms for the three Jones--Scadron transition sectors. In the magnetic channel, the modified dipole remains decisively preferred by AICc, with
\(\chi^2/\nu=0.43\) and Akaike weight \(0.968\). Its fitted parameters are
\begin{align}
A_{M,0}&=3.143\pm0.139,\nonumber\\
\Lambda_M^2&=0.6545\pm0.0319~\mathrm{GeV}^2,\nonumber\\
c_M&=-0.0949\pm0.0086~\mathrm{GeV}^{-2}.
\end{align}
The quartic and cubic \(z\) expansions carry AICc weights \(0.021\) and \(0.011\), respectively, while the pure dipole is strongly disfavored. The magnetic transition form factor therefore remains the most sharply selected empirical normalization of the three sectors.

For the electric channel, the dipole remains selected with
\(\chi^2/\nu=0.55\) and AICc weight \(0.763\), with
\[
A_{E,0}=0.0905\pm0.0192,
\qquad
\Lambda_E^2=0.4489\pm0.0831~\mathrm{GeV}^2.
\]
The modified dipole has a slightly smaller raw \(\chi^2/\nu=0.52\), but the improvement is insufficient to offset the additional parameter in AICc, leaving it with weight \(0.207\). The cubic and quartic \(z\) expansions carry weights \(0.029\) and \(0.002\), respectively.

For the Coulomb channel, the cubic \(z\) expansion is the AICc-selected form, with \(\chi^2/\nu=1.18\) and AICc weight \(0.727\). The fitted coefficients are
\begin{align}
a_{C,0}&=1.174\pm0.234,\nonumber\\
a_{C,1}&=-3.009\pm1.095,\nonumber\\
a_{C,2}&=1.985\pm1.672,\nonumber\\
a_{C,3}&=-0.065\pm0.838.
\end{align}
The quartic \(z\) expansion has AICc weight \(0.163\) and
\(\Delta\mathrm{AICc}=3.00\), while the modified dipole carries weight \(0.110\). The cubic and quartic \(z\) forms are essentially degenerate according to BIC, differing by only \(0.01\), with the quartic form slightly lower. Because AICc is the predefined primary criterion, the cubic form remains the production choice. The Coulomb sector therefore remains more model sensitive than the magnetic channel.

The sensitivity to the alternative within-table covariance prescription is quantified in Section~\ref{sec:discussion} and Table~\ref{tab:model_error_correlation_sensitivity}. The reduced fit qualities should also be interpreted in the context of the adopted uncertainty prescription. The selected magnetic and electric fits have \(\chi^2/\nu=0.43\) and \(0.55\), respectively, while the selected Coulomb fit has \(\chi^2/\nu=1.18\). In the default covariance construction, the published statistical and model uncertainties are combined in quadrature point by point because a complete point-to-point covariance for the quoted model uncertainties is not available. The sub-unity values in the magnetic and electric channels are therefore consistent with a conservative pointwise uncertainty scale and are not used to rescale the uncertainties downward. In particular, the fitted-parameter covariance is not reduced when \(\chi^2/\nu<1\), and the resulting uncertainty sampling is propagated unchanged through the transition-GPD and impact-parameter calculations.

Figure~\ref{fig:form_factor_model_comparison} shows three converted transition-form-factor data sets and corresponding candidate fits. The Jones--Scadron conversion changes the dimensional interpretation and the fitted functional behavior relative to a direct fit of the raw \(\sqrt{\mu{\rm b}}\) multipoles. The magnetic form factor remains dominant, while \(G_E^*\) and \(G_C^*\) are smaller positive quantities in the convention used here, following from the measured negative \(R_{EM}\) and \(R_{SM}\) ratios through the Jones--Scadron relations. The low-\(Q^2\) ratio-sector fits remain separate from these absolute-form-factor fits. In the \(R_{EM}\) sector, the dipole remains the AICc minimum with \(\chi^2/\nu=0.67\) and weight \(0.346\). The cubic \(z\) expansion and Siegert/pion-cloud candidate remain competitive with weights \(0.255\) and \(0.237\), respectively. In the \(R_{SM}\) sector, the modified dipole is the AICc minimum with \(\chi^2/\nu=1.33\) and weight \(0.506\), while the cubic \(z\) expansion is close with weight \(0.411\). The quartic \(z\) form carries weight \(0.083\), and the Siegert/pion-cloud form is strongly disfavored. The ratio-sector analysis remains a low-\(Q^2\) sensitivity diagnostic and does not enter the absolute transition-form-factor normalization used in the tomography.

\section{Form-factor-constrained transition-GPD construction}
\label{sec:gpd}

The Jones--Scadron form-factor fits in Section~\ref{sec:fits} determine the data-constrained momentum-transfer dependence of the \(N\to\Delta\) electromagnetic transition. Now, we introduce a phenomenological \(x\)-dependence while keeping the same imposed form-factor normalization for every longitudinal profile. The electroproduction data determine \(G_M^*(Q^2)\), \(G_E^*(Q^2)\), and \(G_C^*(Q^2)\), but do not determine a unique transition-GPD dependence on longitudinal momentum fraction. The present analysis compares controlled factorized and non-factorized profile families under a common normalization. The channel label is henceforth \(\alpha\in\{M,E,C\}\), with \(Q^2=-t\). The AICc-selected fits determine \(G_\alpha^*(-t)\). The positive-\(x\) normalization \(F_\alpha(t)\) used in the phenomenological GPD profiles is defined explicitly.

The objects constructed here are effective, form-factor-constrained transition-GPD profiles rather than a complete operator-level parametrization of the \(N\to\Delta\) transition GPDs~\cite{Kim:2025CompleteDefinition}. In the standard multipole-GPD convention, the vector transition GPDs satisfy the full-support first-moment relations
\begin{equation}
\int_{-1}^{1}dx\,
H_\alpha(x,\xi,t)
=
2G_\alpha^*(Q^2=-t),\;
\alpha\in\{M,E,C\}.
\label{eq:formal_transition_gpd_moment}
\end{equation}
For the zero-skewness reconstruction used here, the negative-\(x\) region represents the antiquark contribution, so the full first moment can equivalently be represented over \(0<x<1\) by the quark-minus-antiquark, or valence, combination. Because the present ansatz does not introduce an independent antiquark distribution, the modeled positive-\(x\) function is therefore interpreted as an effective valence transition combination. Its imposed normalization is defined as
\begin{equation}
F_\alpha(t)
\equiv
2G_\alpha^*(Q^2=-t).
\label{eq:positive_x_transition_normalization}
\end{equation}
This prescription identifies the positive-\(x\) effective valence combination with the full transition-form-factor first moment. The same valence interpretation is used in the large-\(N_c\) benchmark below. It remains a phenomenological normalization condition, not a derivation of the full operator-level GPD from the electroproduction data.
 The factorized baseline is written as
\begin{equation}
  H_{\alpha}^{\mathrm{fac}}(x,t)=h(x)F_\alpha(t),
  \qquad
  \int_0^1 dx\,h(x)=1 .
  \label{eq:factorized_gpd}
\end{equation}
With this definition, the imposed positive-\(x\) normalization is recovered directly,
\begin{equation}
  \int_0^1 dx\,H_{\alpha}^{\mathrm{fac}}(x,t)=F_\alpha(t).
  \label{eq:factorized_sumrule}
\end{equation}
The longitudinal profile used for the factorized reference is a normalized beta profile,
\begin{equation}
  h(x)=\frac{x^a(1-x)^b}{B(a+1,b+1)} ,
  \label{eq:beta_profile}
\end{equation}
where \(B\) is the Euler beta function. In the baseline profile used here, \(a=0.55\) and \(b=0.35\). These parameters are not fitted to the CLAS multipole data. They define a smooth, normalizable longitudinal reference shape used to test structural sensitivity under the imposed positive-\(x\) form-factor normalization.

The factorized construction is useful because it isolates the effect of the measured form factor. However, it also imposes a strong structural restriction. Because all \(t\)-dependence resides in \(F_\alpha(t)\), the normalized longitudinal shape is independent of momentum transfer. Consequently, after transformation to impact-parameter space, the transverse shape cannot acquire a genuine \(x\)-dependent width from the GPD ansatz itself. This limitation motivates the non-factorized construction below.

\subsection{Non-factorized profile families}

To introduce controlled \(x\)--\(t\) correlations while preserving the same imposed \(F_\alpha(t)\) normalization, each profile family is written in terms of a positive or sign-definite weight \(w_p(x,t)\),
\begin{align}
  H_{\alpha}^{(p)}(x,t)=
  F_\alpha(t)\,
  \frac{w_p(x,t)}
  {N_p(t)} ,\\
  N_p(t)=\int_0^1 dx'\,w_p(x',t).
  \label{eq:nonfactorized_gpd}
\end{align}
This form gives
\begin{equation}
  \int_0^1 dx\,H_{\alpha}^{(p)}(x,t)=F_\alpha(t)
  \quad \mathrm{for\ every\ profile}\ p .
  \label{eq:nonfactorized_sumrule}
\end{equation}
Because \(N_p(t)\) is defined as the integral of the same weight that appears in the numerator, this normalization identity holds algebraically for every admissible profile with finite, nonzero \(N_p(t)\). It is therefore imposed by construction rather than obtained as an independent constraint on the modeled \(x\)-dependence. The profile controls how the fixed transition normalization is distributed over positive \(x\), but it cannot alter that imposed normalization.

\begin{table*}[t]
\centering
\caption{Transition-GPD profile families used in the main comparison. The profile parameters are structural inputs, not fitted quantities, and each profile is normalized at fixed \(t\) to reproduce the imposed \(F_\alpha(t)=2G_\alpha^*(-t)\) normalization.}
\label{tab:gpd_profiles}
{\setlength{\tabcolsep}{4pt}
\begin{tabular}{llll}
\toprule
Profile label & Weight type & Parameters & Role \\
\midrule
factorized\_beta & Eq.~\eqref{eq:beta_profile} & \(a=0.55,\ b=0.35\) & Factorized baseline \\
exp\_xt\_moderate & Eq.~\eqref{eq:exp_xt_profile} & \(a=0.55,\ b=0.35,\ \alpha_x=0.45~\mathrm{GeV}^{-2},\ n=2\) & Central correlated profile \\
regge\_beta & Eq.~\eqref{eq:regge_profile} & \(a_R=0.25,\ b=1.15,\ \alpha'_R=0.35~\mathrm{GeV}^{-2}\) & Regge-like comparison \\
dd\_inspired & Eq.~\eqref{eq:dd_profile} & \(a=0.65,\ b=0.55,\ \alpha_{\mathrm{DD}}=0.55~\mathrm{GeV}^{-2},\ m=2\) & DD-inspired comparison \\
\bottomrule
\end{tabular}
}
\end{table*}

\begin{figure*}[h!]
    \centering
    \includegraphics[width=\textwidth]{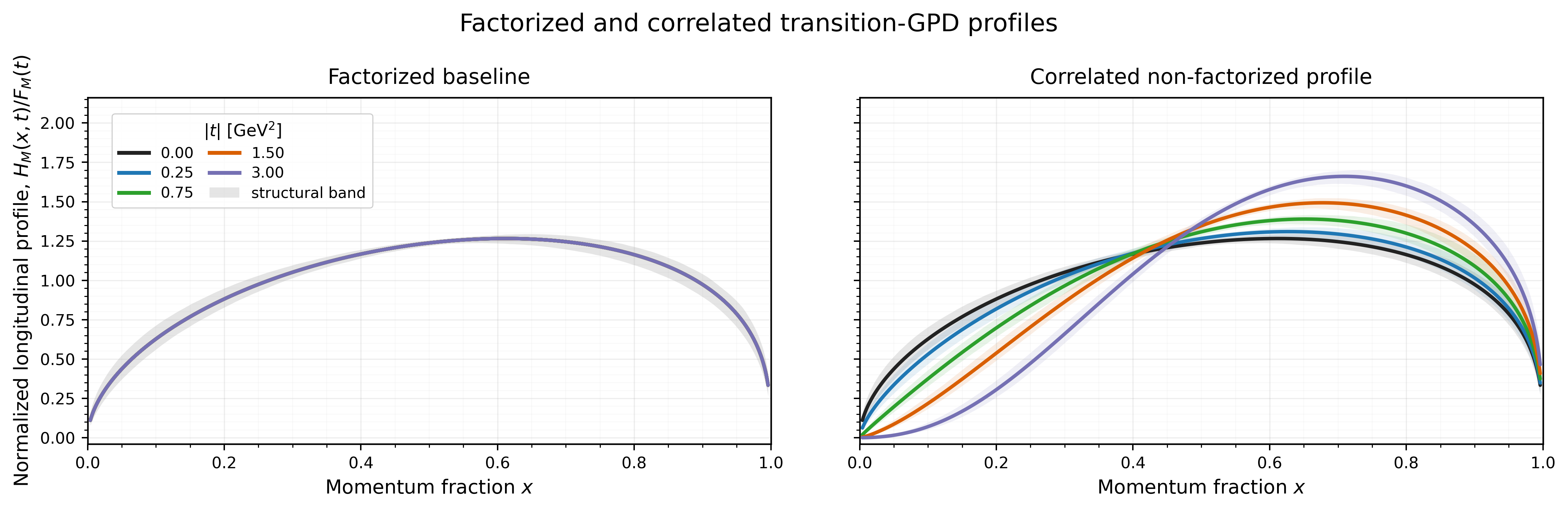}
    \caption{Factorized and correlated transition-GPD profiles in the magnetic channel. Curves show the normalized longitudinal profile \(H_M(x,t)/F_M(t)\) at representative values of \(|t|\). The left panel shows the factorized beta baseline, for which the normalized profile is independent of \(t\). The right panel shows the correlated exponential profile, where the longitudinal shape changes with \(|t|\). The shaded bands show the pointwise 16th--84th percentile ranges from 350 structural replicas. Each continuous parameter of the displayed profile is sampled independently and uniformly over its baseline \(\pm20\%\) interval in Table~\ref{tab:profile_variation_ranges}, with no imposed correlations. Here, \(n=2\) is fixed for the correlated exponential profile. The bands quantify structural sensitivity rather than statistical confidence and do not include uncertainty from the fitted transition form factor.}
    \label{fig:Hxt_profiles}
\end{figure*}

Four profile families are used as the main comparison set. The first is the factorized beta profile of Eq.~\eqref{eq:beta_profile}. The second is a correlated exponential profile,
\begin{equation}
  w_{\mathrm{exp}}(x,t)=
  \frac{x^a(1-x)^b}{B(a+1,b+1)}
  \exp\!\left[t\,\alpha_x(1-x)^n\ln(1/x)\right],
  \label{eq:exp_xt_profile}
\end{equation}
with \(a=0.55\), \(b=0.35\), \(\alpha_x=0.45~\mathrm{GeV}^{-2}\), and \(n=2\). Because \(t<0\) in the spacelike region, this profile changes the longitudinal shape as \(|t|\) increases and produces an explicit correlation between momentum fraction and transverse momentum transfer. The same \(a\) and \(b\) values are used for the factorized and correlated exponential profiles so that their comparison isolates the effect of the \(x\)--\(t\) correlation rather than a change in the baseline longitudinal shape. The third family is a Regge-like beta profile,
\begin{align}
  w_{\mathrm{Regge}}(x,t)&=x^{-\alpha_R(t)}(1-x)^b,
  \label{eq:regge_profile}\\
  \alpha_R(t)&=a_R+\alpha'_R t .
\end{align}
with \(a_R=0.25\), \(b=1.15\), and
\(\alpha'_R=0.35~\mathrm{GeV}^{-2}\). For spacelike \(t<0\), the effective Regge intercept therefore decreases with increasing \(|t|\), making the small-\(x\) behavior progressively less singular. This form provides a \(t\)-dependent small-\(x\) structure analogous to Regge-motivated nucleon-GPD parametrizations~\cite{Radyushkin:1997ki,Goeke:2001tz,Diehl:2003ny}. The fourth family is a double-distribution-inspired profile,
\begin{equation}
  w_{\mathrm{DD}}(x,t)=
  \frac{x^a(1-x)^b}{B(a+1,b+1)}
  \exp\!\left[
  t\,\alpha_{\mathrm{DD}}
  \frac{(1-x)^m}{x+0.08}
  \right],
  \label{eq:dd_profile}
\end{equation}
with \(a=0.65\), \(b=0.55\), \(\alpha_{\mathrm{DD}}=0.55~\mathrm{GeV}^{-2}\), and \(m=2\). This profile is not a full double-distribution construction with skewness dependence. It is a zero-skewness, profile-inspired ansatz designed to mimic a stronger coupling between longitudinal momentum and transverse momentum transfer while retaining the same imposed positive-\(x\) form-factor normalization. 
The profile parameters in Table~\ref{tab:gpd_profiles} are not fitted to the CLAS multipole data. They define structural model families that are compared under the same form-factor normalization. The \(Q^2\)-dependent transition strength is data constrained through the fitted \(G_\alpha^*(Q^2)\), which enters the phenomenological GPD through the imposed normalization \(F_\alpha(t)=2G_\alpha^*(-t)\), whereas the longitudinal profile remains a controlled model assumption. The numerical central values in Table~\ref{tab:gpd_profiles} were chosen phenomenologically to provide smooth, normalizable representative profiles spanning qualitatively different longitudinal and \(x\)--\(t\) structures. They were not tuned to the large-\(N_c\) benchmark. In particular, the baseline exponent \(b=0.35\) is not intended as a valence counting-rule or endpoint determination.

To quantify local sensitivity within each profile family, we use an explicit bounded structural scan. Each continuously varied profile parameter is sampled independently over a symmetric \(\pm20\%\) interval around its central value. No parameter correlations are imposed, and the discrete exponents \(n=2\) and \(m=2\) are held fixed. The uniform sampling is used only as a numerical coverage prescription for the bounded parameter region. It is not interpreted as a Bayesian prior, likelihood-derived distribution, or statistical confidence model for the transition-GPD parameters. The dependence on the chosen range is tested separately using \(\pm10\%\), \(\pm20\%\), and \(\pm30\%\) scans.

\begin{table*}[t]
\centering
\caption{Structural-profile variation prescription. The central values define the four profile families, while the quoted intervals give the baseline \(\pm20\%\) within-family sensitivity range. Continuous parameters are varied independently with uniform coverage of the stated interval and no imposed correlations. The discrete exponents \(n\) and \(m\) are fixed. These ranges define model-sensitivity scans rather than statistical confidence intervals.}
\label{tab:profile_variation_ranges}
\begin{tabular}{lll}
\toprule
Profile & Varied parameters and baseline ranges & Fixed parameters \\
\midrule
Factorized beta &
\(a\in[0.44,0.66],\quad b\in[0.28,0.42]\) &
--- \\

Correlated exponential &
\(a\in[0.44,0.66],\quad b\in[0.28,0.42],\quad
\alpha_x\in[0.36,0.54]~\mathrm{GeV}^{-2}\) &
\(n=2\) \\

Regge-like &
\(a_R\in[0.20,0.30],\quad b\in[0.92,1.38],\quad
\alpha'_R\in[0.28,0.42]~\mathrm{GeV}^{-2}\) &
--- \\

DD-inspired &
\(a\in[0.52,0.78],\quad b\in[0.44,0.66],\quad
\alpha_{\mathrm{DD}}\in[0.44,0.66]~\mathrm{GeV}^{-2}\) &
\(m=2\) \\
\bottomrule
\end{tabular}
\end{table*}

\subsection{Profile comparison in \texorpdfstring{$H(x,t)$}{H(x,t)} space}

The profile dependence is most transparent after dividing out the imposed normalization. Figure~\ref{fig:Hxt_profiles} compares the normalized magnetic-channel quantity \(H_M(x,t)/F_M(t)\) for the factorized baseline and the central correlated exponential profile. The same normalized construction is applied to the electric and Coulomb sectors, but the magnetic channel is shown here because it is dominant and provides the most strongly selected transition-form-factor input. In the factorized case, all \(|t|\) curves coincide because \(F_M(t)\) cancels in the normalized ratio. In the correlated profile, increasing \(|t|\) shifts the longitudinal distribution and changes its width. This behavior is the essential structural difference that later appears as \(x\)-dependent transverse localization after the Hankel transform. The gray bands in Figure~\ref{fig:Hxt_profiles} use the baseline \(\pm20\%\) structural prescription of Table~\ref{tab:profile_variation_ranges}. The continuously varied parameters are sampled independently and uniformly within the stated bounds, with no imposed correlations. The resulting band is therefore a model-sensitivity envelope rather than a statistical confidence interval. Because \(F_M(t)\) cancels in \(H_M(x,t)/F_M(t)\), the displayed variation isolates profile structure rather than uncertainty in the fitted \(G_M^*\).

The numerical implementation of the imposed normalization was checked for all three channels, the four profile families in Table~\ref{tab:gpd_profiles}, and \(t=0,-0.2,-0.5,-1.0,-2.0,-4.0~\mathrm{GeV}^2\). Because the normalization identity follows algebraically from the definition of \(H_\alpha^{(p)}\), this calculation is a quadrature-closure test rather than an independent test of the profile model. Across the \(72\) numerical checks, the largest absolute quadrature residual was \(1.78\times10^{-15}\), with a maximum relative residual of \(5.38\times10^{-16}\). These values show only that the numerical integration reproduces the imposed normalization to floating-point precision. The following section uses these form-factor-normalized transition-GPD profiles as input to the impact-parameter transform. The central physical question then becomes how the different \(x\)--\(t\) profile assumptions map into transverse kernels and localization observables.

\subsection{Large-\(N_c\) magnetic-transition benchmark}

The longitudinal profile is not fixed by the electroproduction form factors alone, but the large-\(N_c\) limit provides an external theoretical reference for the dominant magnetic transition GPD. In the Jones--Scadron normalization used here, Ref.~\cite{Pascalutsa:2007LargeNc} gives the magnetic transition relation
\(H_M=2[G_M^*(0)/\kappa_V](E^u-E^d)\), with
\(\kappa_V=\kappa_p-\kappa_n\). Using the strict leading-\(N_c\) result
\(G_M^*(0)=\kappa_V/\sqrt{2}\) gives
\begin{equation}
H_M(x,\xi,t)
\simeq
\sqrt{2}\,
\left[
E^u(x,\xi,t)-E^d(x,\xi,t)
\right].
\label{eq:largeNc_HM}
\end{equation}
Frankfurt et al.~\cite{Frankfurt:2000LargeNc} use a different isospin normalization for the corresponding transition GPD, \(H_M^{(3)}=(2/\sqrt{3})E^{(3)}\). The differing numerical coefficients therefore reflect convention rather than a physical disagreement. The relation supplies theoretical information on the \(x\)-dependence that is absent from the transition-form-factor normalization itself.

Because the present phenomenological reconstruction models only the positive-\(x\) effective sector, and because the overall normalization of the large-\(N_c\) relation is subject to finite-\(N_c\) and convention effects, the comparison is made at the level of unit-normalized positive-\(x\) shapes. We define
\begin{equation}
\widehat H_M^{(p)}(x,t)
=
\frac{H_M^{(p)}(x,t)}
{\displaystyle\int_0^1 dx'\,
 H_M^{(p)}(x',t)}
\end{equation}
and compare it with
\begin{equation}
\widehat E_V(x,t)
=
\frac{E_v^u(x,0,t)-E_v^d(x,0,t)}
{\displaystyle\int_0^1 dx'\,
 [E_v^u(x',0,t)-E_v^d(x',0,t)]},
\end{equation}
using the zero-skewness valence-GPD parametrization of Diehl, Feldmann, Jakob, and Kroll~\cite{Diehl:2005NucleonGPD}. This benchmark is not included in the fit and is not used to select the transition-profile parameters.

\begin{figure*}[h!]
    \centering
    \includegraphics[width=\textwidth]{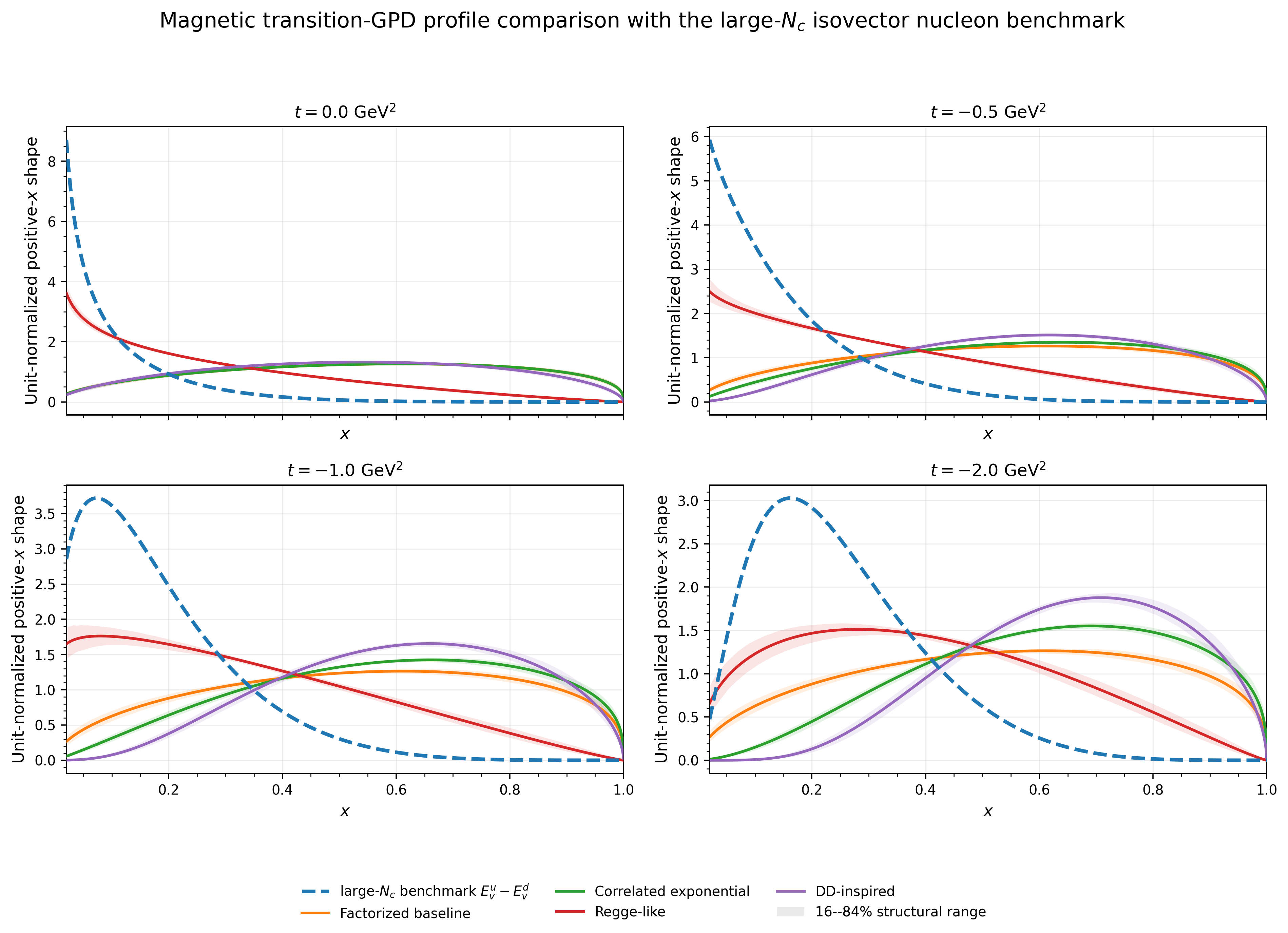}
    \caption{Positive-\(x\) magnetic transition-GPD profile shapes compared with the large-\(N_c\) isovector nucleon benchmark \(E_v^u-E_v^d\) at \(t=0,-0.5,-1.0,\) and \(-2.0~\mathrm{GeV}^2\). All central curves are normalized to unit positive-\(x\) integral, so the comparison tests longitudinal shape rather than overall transition strength. The shaded regions show the pointwise 16th--84th percentile structural-sensitivity ranges obtained from the baseline \(\pm20\%\) within-family parameter scan of Table~\ref{tab:profile_variation_ranges}. They are not statistical confidence intervals. For readability the plotted curves are displayed for \(x\ge0.02\), while the normalization and overlap calculations retain the full positive-\(x\) numerical domain. The nucleon benchmark uses the zero-skewness valence-GPD parametrization of Ref.~\cite{Diehl:2005NucleonGPD}.}
    \label{fig:largeNc_profile_comparison}
\end{figure*}

For a compact quantitative comparison we use the normalized-shape overlap
\begin{equation}
{\cal O}_p(t)
=
\int_0^1 dx\,
\min\!
\left[
\widehat H_M^{(p)}(x,t),
\widehat E_V(x,t)
\right],
\label{eq:largeNc_overlap}
\end{equation}
where \({\cal O}=1\) would denote identical normalized positive-\(x\) shapes. Among the four profile families, the Regge-like ansatz is closest to the large-\(N_c\) benchmark throughout the tested range. Its overlap is \(0.567\), \(0.650\), \(0.672\), and \(0.668\) at \(t=0,-0.5,-1.0,\) and \(-2.0~\mathrm{GeV}^2\), respectively. The corresponding overlaps of the other profile families lie between approximately \(0.22\) and \(0.48\). Within this restricted normalized-shape benchmark, the more strongly small-\(x\)-weighted Regge-like ansatz is qualitatively closest to the large-\(N_c\) nucleon reference. This comparison is not a profile-selection criterion, and none of the four phenomenological families reproduces the benchmark exactly. The within-family structural bands in Figure~\ref{fig:largeNc_profile_comparison} do not alter this qualitative comparison. They show the sensitivity of each transition-profile shape to the prescribed \(\pm20\%\) parameter variation while leaving the external large-\(N_c\) benchmark fixed.

As a separate endpoint stress test, we repeated the factorized and correlated-exponential calculations with \(b=3\), holding all other parameters fixed. At \(t=0\), the normalized strength above \(x=0.8\) decreases from \(0.181\) to \(0.0038\), and the normalized-shape overlap with the large-\(N_c\) benchmark increases from \(0.221\) to \(0.415\). At \(t=-2~\mathrm{GeV}^2\), the overlap increases from \(0.478\) to \(0.878\) for the factorized profile and from \(0.345\) to \(0.688\) for the correlated exponential profile. Thus the longitudinal high-\(x\) shape is strongly endpoint dependent. The primary slope-defined localization is considerably less sensitive. The factorized width is unchanged, while the maximum correlated-profile shift is \(4.94\%\), \(3.41\%\), and \(1.80\%\) in the magnetic, electric, and Coulomb channels, respectively. The strict leading-\(N_c\) normalization underlying Eq.~\eqref{eq:largeNc_HM} is \(G_M^*(0)=\kappa_V/\sqrt{2}\simeq2.62\) in the Jones--Scadron convention used here, compared with \(G_M^*(0)=3.14\) from the empirical transition-form-factor fit. We therefore do not use the large-\(N_c\) relation to replace the empirical normalization. Its role here is deliberately limited to providing an external theoretical reference for the otherwise underconstrained longitudinal shape.

\section{Impact-parameter tomography}
\label{sec:tomography}

The form-factor-normalized transition-GPD profiles constructed in Section~\ref{sec:gpd} can be transformed into transverse coordinate space at fixed longitudinal momentum fraction. In the elastic case, the zero-skewness impact-parameter interpretation of GPDs provides a transverse spatial representation in which the impact parameter \(b\) is conjugate to the transverse momentum transfer \(\Delta\)~\cite{Burkardt:2000za,Burkardt:2002hr,Diehl:2003ny}. For transition matrix elements, the interpretation requires additional care because the initial and final baryon states have different masses, spin quantum numbers, and light-front spin-transition structures. Recent transition-GPD studies emphasize that the corresponding impact-parameter distributions are non-diagonal transition densities, not ordinary diagonal probability densities of either the nucleon or the \(\Delta\)~\cite{Kim:2023QCDAM,Kim:2025TensorPolarized,Kim:2026MultipoleStructure}. The present implementation therefore uses impact-parameter transforms as model-defined spatial diagnostics constrained by the measured transition form factors. Related transverse-density methods have also been applied directly to the nucleon and \(N\to\Delta\) transition form factors~\cite{Vanderhaeghen:2007zz}. The transform is applied to each transition channel and profile family, but the main tomography figure is shown for the magnetic channel because \(M\) supplies the dominant transition amplitude and has the most stable empirical normalization. 

For a profile \(p\), channel \(\alpha\), and Bessel order \(\ell\), the radial impact-parameter transform is defined as
\begin{equation}
\begin{aligned}
q_{\ell,\alpha}^{(p)}(x,b;\Delta_{\max})
&=
\int_0^{\Delta_{\max}} d\Delta\,
\frac{\Delta}{2\pi(\hbar c)^2}
\\
&\quad\times
J_\ell\!\left(\frac{\Delta b}{\hbar c}\right)
H_\alpha^{(p)}(x,-\Delta^2).
\end{aligned}
\label{eq:hankel}
\end{equation}
Here \(\Delta\) is integrated in GeV, \(b\) is expressed in fm, and \(\hbar c=0.1973269804~\mathrm{GeV\,fm}\). The Jones--Scadron transition form factors \(G_M^*\), \(G_E^*\), and \(G_C^*\) are dimensionless, and the modeled transition GPDs \(H_\alpha(x,t)\) are therefore dimensionless. The factor \(1/(\hbar c)^2\) converts the two-dimensional momentum-space transform to units of \(\mathrm{fm}^{-2}\). Consequently \(q_{\ell,\alpha}^{(p)}(x,b)\) has units of \(\mathrm{fm}^{-2}\). The slope-defined transverse width and the finite-window mean radius introduced below have units of fm, while the corresponding finite-window skewness and kurtosis are dimensionless. The order \(\ell=0\) transform is used for the magnetic-channel tomography in this section. Higher-order kernels associated with the Coulomb and electric quadrupole sectors are discussed separately in Section~\ref{sec:multipole_kernels}. For the production reconstruction, the upper integration boundary is tied to the largest measured absolute transition-form-factor point,
\[
\Delta_{\max}=\sqrt{Q^2_{\max,\mathrm{data}}}
             =\sqrt{6}~\mathrm{GeV}
             =2.44949~\mathrm{GeV}.
\]
The production grid uses \(N_x=70\), \(N_b=70\), and \(N_\Delta=280\). The \(x\) grid excludes the exact endpoints, the \(b\) grid covers \(0.01\leq b\leq2.5~\mathrm{fm}\), and the \(\Delta\) integral is evaluated by direct quadrature. The resulting object is therefore a band-limited transverse kernel whose upper momentum-transfer boundary is fixed by the largest measured absolute transition-form-factor point, rather than an infinite-momentum-transfer Fourier reconstruction. For compactness, the explicit \(\Delta_{\max}\) argument is suppressed in the notation below.

\subsection{Magnetic-channel tomography}

\begin{figure*}[h!]
    \centering
    \includegraphics[width=\textwidth]{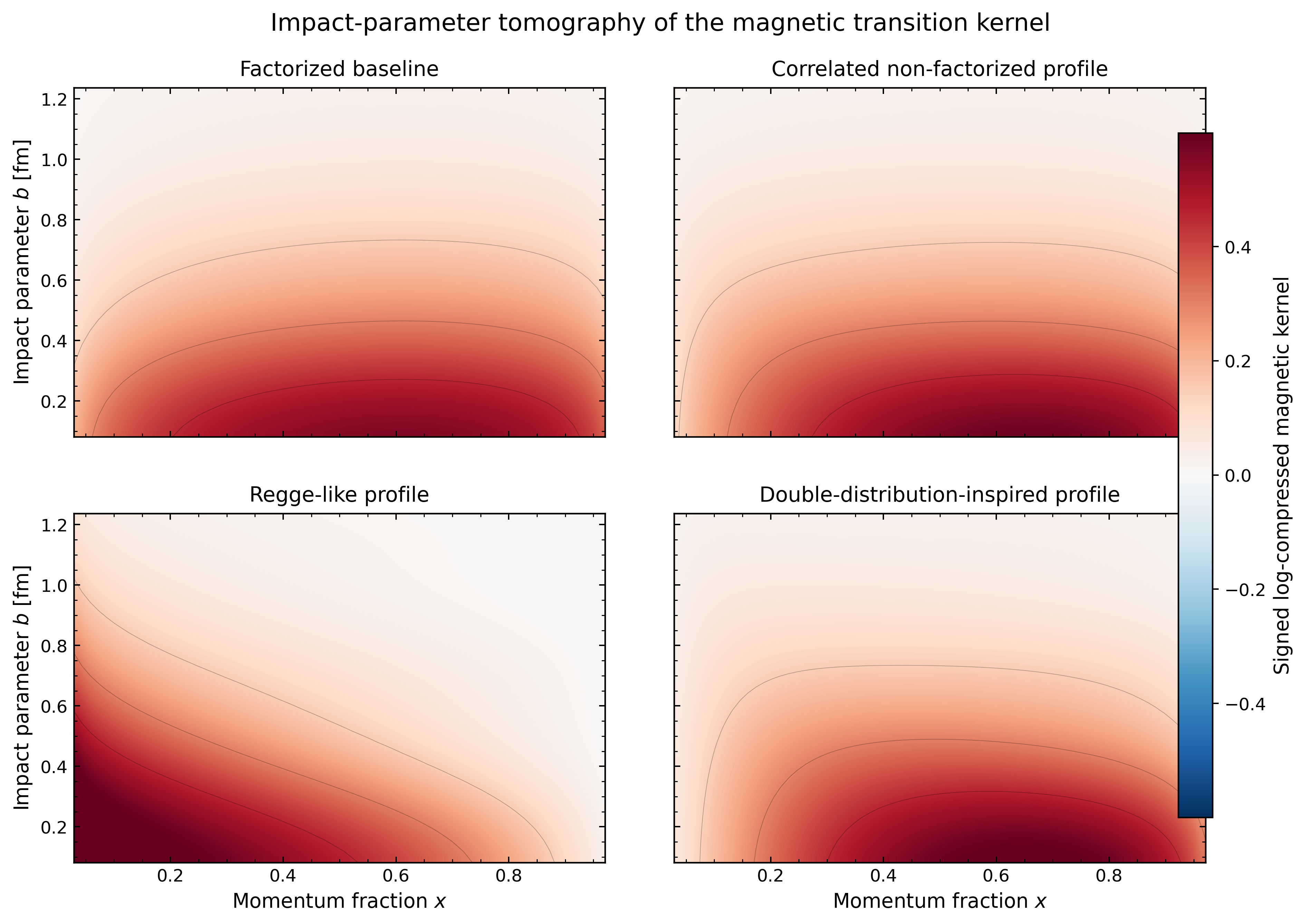}
    \caption{Impact-parameter representation of the magnetic transition kernel. The panels show the signed-log-compressed \(J_0\) Hankel transform \(q_M^{(p)}(x,b)\) for the factorized beta, correlated non-factorized, Regge-like, and double-distribution-inspired profiles. The same normalization derived from the AICc-selected \(G_M^*\) fit is used in all panels. Differences among panels therefore reflect the assumed \(x\)--\(t\) profile structure.}
    \label{fig:qxb_tomography}
\end{figure*}

The magnetic-channel transverse kernel shown in Figure~\ref{fig:qxb_tomography} is defined as
\begin{equation}
  q_M^{(p)}(x,b)\equiv q_{0,M}^{(p)}(x,b),
  \label{eq:magnetic_qxb}
\end{equation}
computed from the AICc-selected \(G_M^*\) fit, with \(F_M(t)=2G_M^*(-t)\), and each of the four profile families in Table~\ref{tab:gpd_profiles}. The figure compares the factorized baseline, the central correlated non-factorized profile, the Regge-like profile, and the double-distribution-inspired profile. The same imposed magnetic normalization \(F_M(t)=2G_M^*(-t)\) is used in all four panels, so differences among the maps arise from the profile-dependent \(x\)--\(t\) structure rather than from changes to the measured transition strength.

The factorized baseline produces a transverse kernel whose \(x\)-dependence is inherited only from the fixed longitudinal profile. Its largest-magnitude region remains concentrated at small \(b\) with a broad maximum near the peak of the beta profile. The correlated non-factorized profile shifts the strength toward larger \(x\) as the transverse momentum transfer increases, producing a visibly different localization pattern even though the same imposed magnetic normalization is recovered after integration over \(x\). The Regge-like profile gives the most pronounced small-\(x\) enhancement and a broader low-\(x\) transverse structure. The double-distribution-inspired profile lies between the correlated exponential and Regge-like cases, with a strong but smoother redistribution of magnetic transition strength in the \(x\)--\(b\) plane. This comparison illustrates the main physical role of non-factorization in the present analysis. The fitted \(G_M^*(Q^2)\), through the imposed normalization \(F_M(t)=2G_M^*(-t)\), fixes the integrated transition strength, but it does not determine how that strength is distributed over \(x\). Once an \(x\)--\(t\) correlation is introduced, different longitudinal momentum regions can acquire different transverse localization scales. The maps in Figure~\ref{fig:qxb_tomography} show the spatial consequences of profile structure under a common empirical normalization.

\subsection{Numerical implementation and display}

The maps in Figure~\ref{fig:qxb_tomography} are displayed over the approximate window \(0.03\leq x\leq0.97\) and \(0.08\leq b\leq1.24~\mathrm{fm}\), while the transform is evaluated on the full production grid. For visualization, the plotted values use the signed compression \(q_c=\operatorname{sgn}(q)\log_{10}[1+|q|/q_0]\) with \(q_0=3.04~\mathrm{fm}^{-2}\). This compression does not enter any localization observable. A separate cutoff-sensitivity study was performed to determine whether extending the selected transition-form-factor fits beyond the measured region can affect the transverse reconstruction. The test compares the unnormalized kernels obtained with \(\Delta_{\max}=\sqrt{6},\,3,\,4,\,5,\,6.5,\,8,\) and \(10~\mathrm{GeV}\) at representative momentum fractions \(x=0.20,\ 0.60,\) and \(0.85\). The transition-GPD profile is held fixed to the central correlated exponential profile throughout this diagnostic so that the effect of the transform boundary is isolated from the separate profile-family dependence studied elsewhere in the manuscript. With the two-pion \(z\)-mapping of Eq.~\eqref{eq:zvariable}, the AICc-selected cubic \(G_C^*\) fit gives \(G_C^*(6~\mathrm{GeV}^2)=0.003816\) and \(G_C^*(64~\mathrm{GeV}^2)=0.04144\) when extrapolated to the \(\Delta_{\max}=8~\mathrm{GeV}\) boundary, an increase by a factor of approximately \(10.9\) relative to the last measured point. The corresponding 8-GeV calculation changes the finite-window rms diagnostic relative to the data-domain-limited result by as much as approximately \(6.3\%\) in the magnetic channel, \(5.9\%\) in the electric channel, and \(21.3\%\) in the Coulomb channel. For the central correlated Coulomb profile, the unnormalized 8-GeV kernel is within \(5\%\) of the data-domain-limited result only for \(b\gtrsim0.39~\mathrm{fm}\) at \(x=0.60\) and \(b\gtrsim1.93~\mathrm{fm}\) at \(x=0.85\). The production transform therefore remains restricted to \(\Delta_{\max}=\sqrt6~\mathrm{GeV}\), while the higher-cutoff calculations are retained only as diagnostics of extrapolation sensitivity.

\begin{figure*}[h!]
    \centering
    \includegraphics[width=\textwidth]{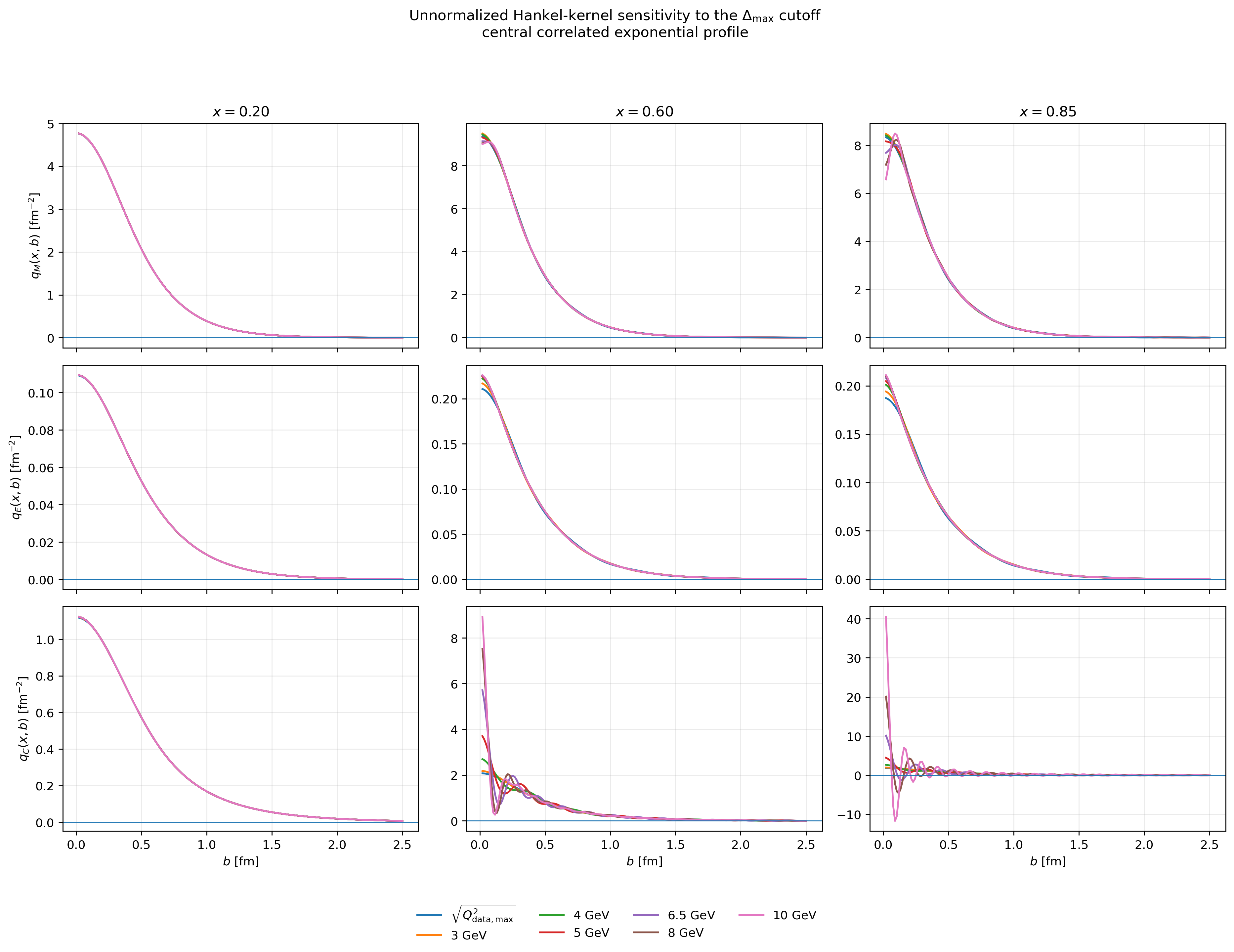}
    \caption{Cutoff sensitivity of the unnormalized order-zero transition kernels with the transition-GPD profile fixed to the central correlated exponential form. Rows show the magnetic, electric, and Coulomb channels and columns show \(x=0.20,\ 0.60,\) and \(0.85\). Holding the profile fixed isolates sensitivity to the transform boundary from the separate structural-profile dependence. The blue curve uses the data-supported boundary \(\Delta_{\max}=\sqrt6~\mathrm{GeV}\). The remaining curves extend the selected empirical form-factor fits beyond that boundary only as extrapolation diagnostics. The Coulomb kernel shows the strongest cutoff dependence when its selected cubic-\(z\) fit is extrapolated outside the measured region. All production results use only \(\Delta_{\max}=\sqrt6~\mathrm{GeV}\).}
    \label{fig:cutoff_sensitivity}
\end{figure*}

Numerical quadrature convergence was then tested independently with the physical cutoff held fixed at \(\Delta_{\max}=\sqrt6~\mathrm{GeV}\). Using the central correlated profile and all three \(M\), \(E\), and \(C\) channels, calculations with \(N_\Delta=140,\ 280,\) and \(560\) were compared over \(0.02\leq b\leq2.5~\mathrm{fm}\) at \(x=0.20,\ 0.60,\) and \(0.85\). Relative to the \(N_\Delta=560\) reference, the production choice \(N_\Delta=280\) gives a maximum scaled kernel difference of \(3.66\times10^{-5}\), or \(0.0037\%\). The adopted production grid is therefore numerically converged to substantially better than the percent level once the physical transform domain is held fixed. Figure~\ref{fig:cutoff_sensitivity} displays the unnormalized cutoff dependence directly.

\section{Slope-defined localization and finite-window shape diagnostics}
\label{sec:localization}

The band-limited transverse kernels of Section~\ref{sec:tomography} provide a direct coordinate-space representation of the modeled transition strength, but a sharp upper boundary in transverse momentum necessarily produces weak oscillatory large-\(b\) ringing. This matters for radial moments because powers of \(b\) can amplify very small long-range oscillations. A dedicated \(b_{\max}\) study shows that the \(|q|\)-weighted moments do not approach stable infinite-range values when the radial integration boundary is increased. The primary localization scale is therefore defined directly from the small-\(Q^2\) slope of the transition GPD, while mean-radius and higher-shape information are retained separately as finite-window diagnostics of the resolved kernel core.

For channel \(\alpha\) and profile family \(p\), the primary transverse localization scale is defined by
\begin{equation}
\begin{aligned}
\left[b_{{\rm rms},\alpha,p}^{\rm slope}(x)\right]^2
&=
-4(\hbar c)^2
\frac{1}{H_\alpha^{(p)}(x,0)}
\\
&\quad\times
\left.
\frac{\partial H_\alpha^{(p)}(x,-Q^2)}
{\partial Q^2}
\right|_{Q^2=0}.
\end{aligned}
\label{eq:slope_brms}
\end{equation}
This expression is the familiar small-\(|t|\) transverse-width relation in the diagonal case. Here it is used more conservatively as a slope-defined localization scale of the modeled non-diagonal transition kernel rather than as the rms radius of a positive probability density. Accordingly, \(b_{\rm rms}^{\rm slope}(x)\) should be read as the transverse width associated with the local momentum-transfer slope of the modeled form-factor-constrained transition amplitude at fixed \(x\). It is not the geometric radius of either the initial nucleon or the final \(\Delta\), nor is it the rms radius of a complete polarized \(N\to\Delta\) light-front charge or current density.

Using the normalized construction of Eq.~\eqref{eq:nonfactorized_gpd}, the slope-defined width can be written directly as
\begin{equation}
\begin{aligned}
\left[b_{{\rm rms},\alpha,p}^{\rm slope}(x)\right]^2
&=
-4(\hbar c)^2
\frac{G_\alpha^{*\prime}(0)}
     {G_\alpha^*(0)}
\\
&\quad
+4(\hbar c)^2
\left[
g_p(x)-\langle g_p\rangle_p
\right],
\end{aligned}
\label{eq:slope_closed_form}
\end{equation}
where
\begin{equation}
\begin{aligned}
g_p(x)
&\equiv
-\left.
\frac{\partial\ln w_p(x,-Q^2)}
{\partial Q^2}
\right|_{Q^2=0},
\\
\langle g_p\rangle_p
&=
\frac{\displaystyle\int_0^1 dx\,
w_p(x,0)\,g_p(x)}
{\displaystyle\int_0^1 dx\,
w_p(x,0)}.
\end{aligned}
\label{eq:profile_slope_function}
\end{equation}
For the correlated exponential, Regge-like, and DD-inspired profiles,
\begin{align}
    g_p(x)=
    \alpha_x(1-x)^n\ln(1/x),\\g_p(x)=
    \alpha'_R\ln(1/x),\\g_p(x)=
    \alpha_{\rm DD}\frac{(1-x)^m}{x+0.08},
\end{align}
respectively, while \(g_p(x)=0\) for the factorized profile. With the central parameters of Table~\ref{tab:gpd_profiles}, the corresponding averages are
\(\langle g_p\rangle_p=0.177,\ 0.689,\) and \(0.529~\mathrm{GeV}^{-2}\) for the correlated exponential, Regge-like, and DD-inspired profiles. Equation~\eqref{eq:slope_closed_form} makes the separation explicit. The transition-form-factor contribution is channel dependent but independent of \(x\), whereas the \(x\)-dependent term is determined entirely by the assumed profile \(w_p\). Because the same profile functions are used in the magnetic, electric, and Coulomb sectors, the profile-induced \(x\) dependence of \([b_{\rm rms}^{\rm slope}(x)]^2\) is common to the three channels. Their form factors provide different constant offsets.

The same decomposition also clarifies the \(x\to1\) limit. For the central magnetic fit, the profile averages correspond to approximately \(6\%\), \(22\%\), and \(17\%\) of the magnitude of the form-factor logarithmic slope for the correlated exponential, Regge-like, and DD-inspired profiles, respectively. Because \(g_p(x)\to0\) as \(x\to1\) for all three forms, their slope-defined widths approach finite values of \(0.681\), \(0.619\), and \(0.639~\mathrm{fm}\), respectively. In diagonal GPDs, a vanishing transverse width as \(x\to1\) follows from the transverse-center-of-momentum interpretation. The present \(N\to\Delta\) kernel is instead a non-diagonal transition matrix element and is not assigned a probability-density interpretation, so we do not impose the diagonal \(x\to1\) condition as an additional constraint. If it were imposed phenomenologically, \(-G_M^{*\prime}(0)/G_M^*(0)=\langle g_p\rangle_p\) would require approximately \(\alpha_x=8.0\), \(\alpha'_R=1.6\), and \(\alpha_{\rm DD}=3.3~\mathrm{GeV}^{-2}\), well outside the structural ranges explored here. The production calculation evaluates Eq.~\eqref{eq:slope_closed_form} analytically. As a numerical cross-check, a cubic local fit gives Coulomb correlated-profile widths lower by approximately \(0.006~\mathrm{fm}\) over \(x=0.45\)--\(0.55\), consistent with the finite-fitting-window sensitivity of that approximation. Because the absolute Jones--Scadron input grid begins at \(Q^2=0.3~\mathrm{GeV}^2\), the form-factor contribution at \(Q^2=0\) remains an extrapolative property of the selected empirical fit. The quoted fit-only bands propagate parameter covariance within that selected functional form and do not include functional-form or model-selection uncertainty.

The associated low-\(Q^2\) functional-form dependence was quantified by evaluating the factorized slope width for the alternative form-factor fits carrying non-negligible AICc support. In the electric channel, the dipole, modified-dipole, and cubic-\(z\) candidates give \(b_{\rm rms}^{\rm slope}=0.833,\ 0.732,\) and \(0.890~\mathrm{fm}\), respectively. In the Coulomb channel, the cubic-\(z\), quartic-\(z\), and modified-dipole candidates give \(1.132,\ 2.914,\) and \(0.830~\mathrm{fm}\). The large quartic-\(z\) value accompanies a continuation in which \(G_C^*(Q^2)\) changes sign between \(Q^2=0.3~\mathrm{GeV}^2\) and the photon point. As an independent diagnostic, comparison with the external low-\(Q^2\) \(R_{SM}\) points gives maximum absolute residuals of approximately \(1.25\sigma\), \(2.90\sigma\), and \(0.91\sigma\) for the cubic-\(z\), quartic-\(z\), and modified-dipole Coulomb continuations, respectively. These ratio-sector points are not included in the absolute-form-factor fit or in the AICc selection.

Mean-radius and higher-shape information are retained as secondary diagnostics of the resolved central part of the band-limited kernel. For this purpose a fixed analysis window $b_{\rm win}=2.5~\mathrm{fm}$ is used, matching the documented production \(b\) domain. At fixed \(x\), define the normalized magnitude-weighted radial distribution
\begin{equation}
\begin{aligned}
P_{\alpha,p}^{\rm win}(b|x)
&=
\frac{
b\,\left|q_\alpha^{(p)}(x,b)\right|
}{
\displaystyle
\int_0^{b_{\rm win}} db'\,
b'\,\left|q_\alpha^{(p)}(x,b')\right|
},
\end{aligned}
\label{eq:windowed_radial_distribution}
\end{equation}
for $0\le b\le b_{\rm win}$. Its finite-window raw moments are
\begin{equation}
\mu_{n,\alpha,p}^{\rm win}(x)
=
\int_0^{b_{\rm win}}
db\,
b^n
P_{\alpha,p}^{\rm win}(b|x).
\label{eq:windowed_raw_moments}
\end{equation}
The windowed mean radius is therefore
\begin{equation}
\langle b\rangle_{\alpha,p}^{\rm win}(x)
=
\mu_{1,\alpha,p}^{\rm win}(x).
\label{eq:windowed_mean}
\end{equation}
These finite-window quantities are not identified with moments of the full \(b\in[0,\infty)\) transform. They are deliberately restricted core-shape diagnostics that summarize where the numerically resolved band-limited kernel carries its dominant magnitude within the documented analysis window. In particular, \(\langle b\rangle^{\rm win}(x)\) denotes the magnitude-weighted mean transverse coordinate of the resolved kernel core within the specified window. It is not a physical hadron radius or a moment of a complete transition density.

\begin{figure*}[h!]
    \centering
    \includegraphics[width=\textwidth]{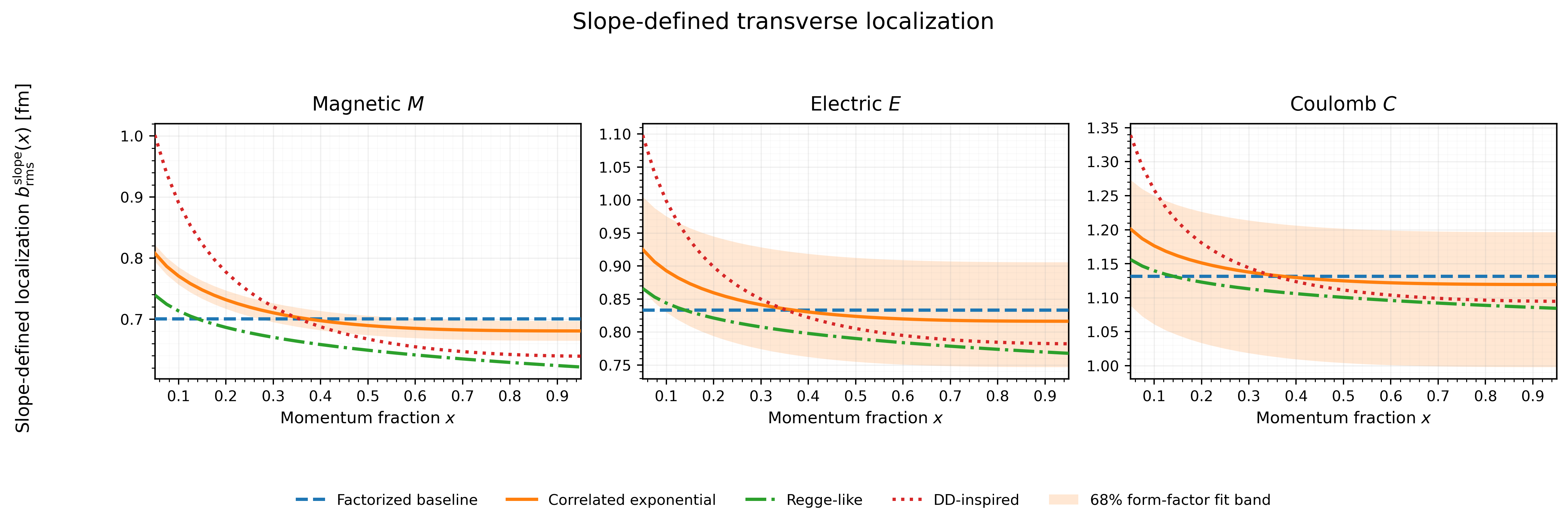}
    \caption{Slope-defined transverse localization scale \(b_{\rm rms}^{\rm slope}(x)\) for the magnetic, electric, and Coulomb transition channels. Curves show the four central transition-GPD profile families. The shaded region in each panel is the 16th--84th percentile uncertainty from the selected transition-form-factor fit, evaluated with \(N_{\rm rep}=5000\) while the central correlated exponential profile is held fixed. The band therefore represents empirical form-factor-fit uncertainty and is not combined with the discrete spread among profile families.}
    \label{fig:localization_all_channels}
\end{figure*}

Figure~\ref{fig:localization_all_channels} shows the primary slope-defined localization scale for all three transition channels. The four central curves display the discrete dependence on the assumed transition-GPD profile, while the shaded band in each panel shows the separate \(68\%\) selected-form-factor fit uncertainty with the central correlated exponential profile held fixed. The figure therefore separates empirical form-factor uncertainty from structural profile-family dependence rather than combining them into a single uncertainty measure. Direct inspection of the signed Coulomb kernels confirms smooth evolution through the mid-\(x\) regions in which the localization scale is most profile sensitive. For the correlated exponential profile at \(x=0.45,\ 0.50,\) and \(0.55\), the analytic slope-defined widths are \(1.1269,\ 1.1248,\) and \(1.1232~\mathrm{fm}\). For the DD-inspired profile at \(x=0.60,\ 0.65,\) and \(0.70\), they are \(1.1038,\ 1.1012,\) and \(1.0991~\mathrm{fm}\). Both the underlying kernels and the slope-defined localization scale therefore vary smoothly across these representative regions. The complementary finite-window quantities \(\sqrt{\mu_2^{\rm win}}\) and \(\langle b\rangle^{\rm win}\) show the same qualitative distinction between the factorized and non-factorized constructions, but they are retained only as secondary diagnostics of the resolved kernel core over \(b\le2.5~\mathrm{fm}\). Their fit and within-family structural sensitivities are quantified separately in Table~\ref{tab:uncertainty_decomposition_localization}. These quantities are not used as global localization radii. Equation~\eqref{eq:slope_closed_form} also clarifies the channel comparison. For a fixed profile family, the \(x\)-dependent contribution to \([b_{\rm rms}^{\rm slope}]^2\) is identical in the magnetic, electric, and Coulomb channels, while the fitted form factors determine different \(x\)-independent offsets. The form-factor data therefore constrain the channel-dependent slope scale but not the modeled \(x\) dependence. Complementary channel-dependent structural information is instead provided by the explicit finite-window kernels and radial-shape diagnostics.

\subsection{Relation to the tomography maps}

The slope-defined width and the finite-window shape diagnostics summarize different aspects of the same modeled transition kernel. The quantity \(b_{\rm rms}^{\rm slope}(x)\) is determined locally by the small-\(Q^2\) behavior of \(H_\alpha(x,-Q^2)\) and is therefore insensitive to the artificial long-range ringing generated by a sharp band limit. The finite-window mean radius and standardized shape moments instead describe the resolved core of the explicit \(q_\alpha(x,b)\) kernel over \(b\le2.5~\mathrm{fm}\). Both constructions show that factorization suppresses nontrivial \(x\)-dependent spatial evolution and that the non-factorized profiles produce distinct localization patterns under the same empirical normalization. The two classes of diagnostics are therefore complementary. The slope quantity supplies the robust primary width, while the finite-window quantities characterize the detailed radial shape of the reconstructed core.

To determine whether the magnitude-weighted radial moments define stable global observables, the radial boundary was varied in representative Coulomb-profile regions. Using \(b_{\max}=2.5,\ 3.5,\ 5.0,\ 7.5,\) and \(10.0~\mathrm{fm}\), the correlated exponential profile at \(x=0.50\) gives a magnitude-weighted rms value increasing from approximately \(0.535~\mathrm{fm}\) to \(1.231~\mathrm{fm}\), while the DD-inspired profile at \(x=0.65\) increases from approximately \(0.517~\mathrm{fm}\) to \(1.422~\mathrm{fm}\). The continued growth demonstrates that these quantities are not converged infinite-range moments of the sharply band-limited kernel. For this check we define the cancellation ratio as
\begin{equation}
R_{\rm canc}(x;b_{\max})=\frac{
\left|\int_0^{b_{\max}}db\,2\pi b\,q_\alpha(x,b)\right|}{
\int_0^{b_{\max}}db\,2\pi b\,|q_\alpha(x,b)|}.
\end{equation}
A value \(R_{\rm canc}=1\) corresponds to no sign cancellation. The ratio remains close to unity in the \(b\le2.5~\mathrm{fm}\) core window, but it is not assumed to be identically unity across all channels, profiles, and \(x\) values because weak cutoff-induced sign changes can occur. The absolute value in Eq.~\eqref{eq:windowed_radial_distribution} is therefore retained to define a uniform nonnegative finite-window shape weight. When the radial boundary is enlarged, the signed second moment also becomes sensitive to the oscillatory band-limit tail. These tests motivate the slope-defined width as the primary localization observable and restrict the magnitude-weighted moments to the documented \(b_{\rm win}=2.5~\mathrm{fm}\) core window.

In the Coulomb channel, the non-probabilistic envelope across the four central profile families has a median half-width of approximately \(0.020~\mathrm{fm}\), compared with a median \(68\%\) selected-\(G_C^*\) fit half-width of approximately \(0.099~\mathrm{fm}\) for the central correlated profile. Their ratio is approximately \(0.20\), so the empirical form-factor uncertainty exceeds the discrete profile-family spread over most of the evaluated \(x\) range, although profile dependence remains visible at low \(x\). This separation is already displayed in Figure~\ref{fig:localization_all_channels}, where the four central profile curves are shown together with the fit-only uncertainty band.

\section{Multipole-resolved radial kernels}
\label{sec:multipole_kernels}

Sections~\ref{sec:tomography} and~\ref{sec:localization} used the order-zero impact-parameter transform to study the \(x\)-dependent transverse localization of the transition strength. That construction is most directly associated with the magnetic channel, which supplies the dominant transition amplitude and provides the cleanest scalar tomography observable. The \(N\to\Delta(1232)\) transition, however, is not a single-multipole process. Recent formal work on \(N\to\Delta\) transition GPDs provides the appropriate context for interpreting multipole structure in non-diagonal light-front transition matrix elements~\cite{Kim:2025CompleteDefinition,Kim:2026MultipoleStructure}. 

The purpose of this section is therefore deliberately restricted. It separates the \(J_0\), \(J_1\), and \(J_2\) radial Bessel-kernel content associated here with the \(M\), \(C\), and \(E\) sectors under the same form-factor-normalized transition-GPD construction used throughout the paper. These kernels are not intended to represent a complete physical transverse charge or current density with all angular factors, spin projections, and convention-dependent light-front current normalizations included. A full polarized \(N\to\Delta\) transverse-density analysis requires fixing the complete light-front spin-transition basis, angular dependence, and current convention, as well as its mapping to the electroproduction multipoles~\cite{Jones:1972ky,Pascalutsa:2006up,Vanderhaeghen:2007zz,Kim:2025CompleteDefinition,Kim:2026MultipoleStructure}. The present analysis instead keeps the comparison at the level of testable radial-kernel diagnostics. This convention is the same one used in Section~\ref{sec:localization} where \(b_{\rm rms}^{\rm slope}\) is a slope-defined width of the modeled transition kernel, and \(\langle b\rangle^{\rm win}\), skewness, and kurtosis characterize only its resolved finite-window radial shape. None of these quantities is identified with the geometric radius of either hadron or with a complete polarized transition-density radius.

\begin{figure*}[h!]
    \centering
    \includegraphics[width=\textwidth]{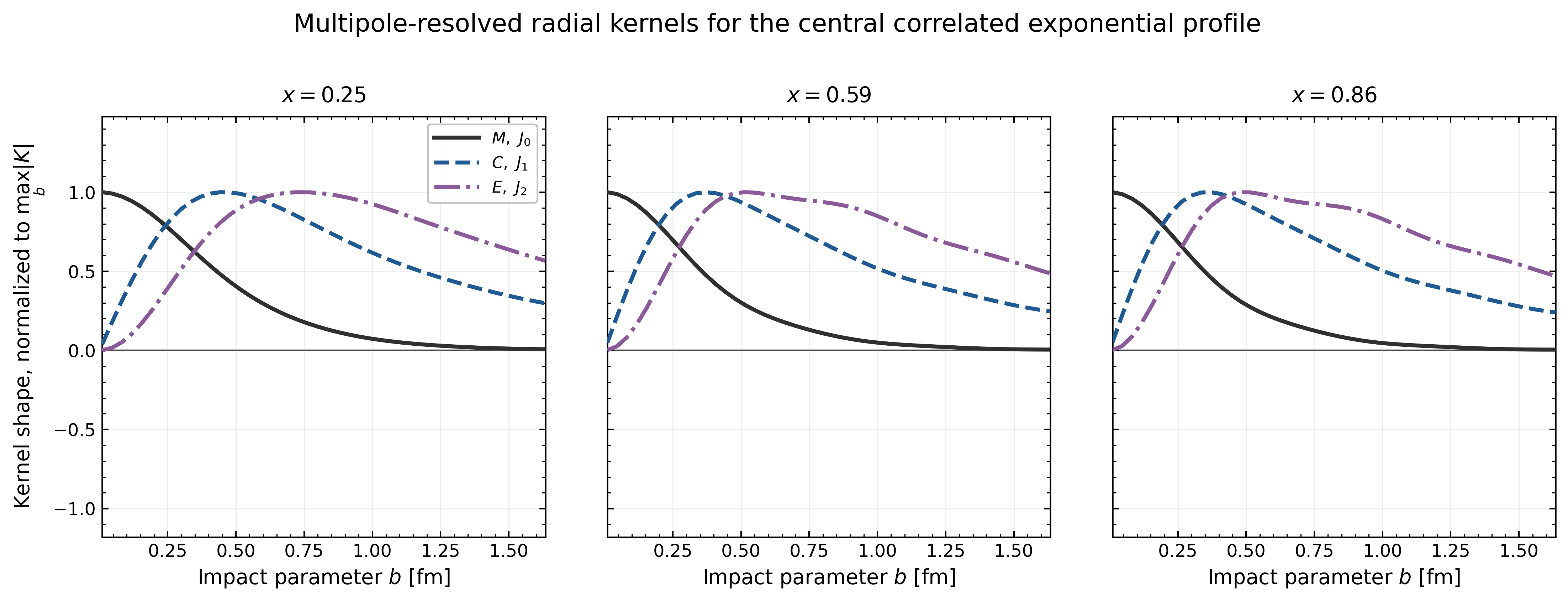}
    \caption{Multipole-resolved radial kernels for the central correlated exponential profile. The panels show representative momentum fractions \(x=0.25\), \(0.59\), and \(0.86\), corresponding to the nearest numerical grid points used in the transform. Curves show the shape-normalized magnetic \(M,J_0\), Coulomb \(C,J_1\), and electric \(E,J_2\) radial kernels. Each curve is divided by its own maximum absolute value at fixed \(x\), so the figure compares radial shape rather than absolute multipole strength.}
    \label{fig:multipole_kernels}
\end{figure*}

For the central correlated non-factorized profile, the three displayed kernels are defined as
\begin{align}
  K_M(x,b) \equiv q_{0,M}^{(p)}(x,b),\\
  K_C(x,b) \equiv q_{1,C}^{(p)}(x,b),\\
  K_E(x,b) \equiv q_{2,E}^{(p)}(x,b).
  \label{eq:multipole_kernels}
\end{align}
In this diagnostic representation, the magnetic sector is shown with the \(J_0\) monopole-like radial transform, the Coulomb sector with the \(J_1\) radial transform, and the electric quadrupole sector with the \(J_2\) radial transform. This assignment should be interpreted as a kernel decomposition, not as a complete angular density. The magnetic kernel gives the scalar radial component already used for the tomography maps, while the \(J_1\) and \(J_2\) kernels isolate how the Coulomb and electric quadrupole sectors populate transverse radius under their corresponding Bessel orders. Because the three multipole sectors have very different absolute magnitudes, Figure~\ref{fig:multipole_kernels} uses a shape-normalized display. At each shown value of \(x\), each kernel is divided by its own maximum absolute value over the displayed \(b\) range,
\begin{equation}
  \widehat K_\alpha(x,b)=
  \frac{K_\alpha(x,b)}
  {\max_b |K_\alpha(x,b)|},
  \qquad
  \alpha\in\{M,C,E\}.
  \label{eq:kernel_normalization}
\end{equation}
This normalization makes the radial shapes of the subleading \(C\) and \(E\) kernels visible on the same axes as the dominant magnetic kernel. It removes the relative absolute normalization among the multipoles, so Figure~\ref{fig:multipole_kernels} should be read as a comparison of radial structure rather than a comparison of transition strength.

The effect of the omitted cross-channel covariance was tested by repeating the form-factor extraction in a joint \(M/E/C\) fit using the complete covariance block obtained from the common replica ensemble. Over representative momentum transfers between \(Q^2=0\) and \(6~\mathrm{GeV}^2\), the resulting \(G_E^*/G_M^*\) and \(G_C^*/G_M^*\) ratios differ from the independently fitted values by at most \(1.16\%\) and \(7.27\%\), respectively. When propagated to the kernels shown here, the raw peak magnitudes change by no more than \(0.05\%\) in \(M\), \(1.30\%\) in \(C\), and \(0.71\%\) in \(E\). After the shape normalization of Eq.~\eqref{eq:kernel_normalization}, the maximum pointwise differences are \(0.11\%\), \(1.45\%\), and \(0.50\%\), respectively. The cross-channel covariance therefore has its largest relative-normalization effect in the Coulomb sector but does not change the radial-shape interpretation of Figure~\ref{fig:multipole_kernels}.

Figure~\ref{fig:multipole_kernels} shows the shape-normalized kernel comparison at three representative momentum fractions. The actual grid values used in the figure are \(x=0.246\), \(0.594\), and \(0.855\). The magnetic \(J_0\) kernel is positive for all three slices, reaches its maximum at the smallest displayed impact parameter, \(b=0.01~\mathrm{fm}\), and then decreases monotonically with \(b\). Its normalized radial shape becomes modestly more concentrated toward small \(b\) as \(x\) increases. The Coulomb \(J_1\) kernel has a qualitatively different radial structure. It is suppressed near the origin, rises to a positive finite-\(b\) maximum, and then decreases smoothly toward larger transverse radius. The maximum occurs at approximately \(b=0.44~\mathrm{fm}\) for \(x=0.246\) and \(b=0.37~\mathrm{fm}\) for \(x=0.594\) and \(0.855\). The electric \(J_2\) kernel is more strongly suppressed near the origin and reaches a broader maximum at larger \(b\), at approximately \(0.73\), \(0.52\), and \(0.48~\mathrm{fm}\) for the three displayed momentum fractions, respectively. These peak positions and their inward shift with \(x\) refer specifically to the band-limited production kernels evaluated at \(\Delta_{\max}=\sqrt6~\mathrm{GeV}\).

None of the three displayed shape-normalized kernels changes sign over the plotted \(b\) interval. Because each curve is normalized by its own maximum absolute value at fixed \(x\), Figure~\ref{fig:multipole_kernels} compares radial shape rather than absolute transition strength. The distinct \(J_0\), \(J_1\), and \(J_2\) profiles therefore show how the magnetic, Coulomb, and electric sectors populate transverse radius differently within the same central correlated exponential transition-GPD profile. They should be interpreted as multipole-resolved radial kernels rather than as separate complete physical transverse densities. The normalized comparison highlights why the multipole decomposition is useful. The magnetic channel controls the dominant scalar tomography, but the subleading Coulomb and electric sectors carry different radial information. These differences would be largely obscured in an unnormalized comparison because the magnetic contribution is much larger. By separating the Bessel orders, the kernel representation provides a compact diagnostic of deformation-sensitive transverse structure while preserving the convention-safe interpretation of the previous sections.

\section{Discussion}
\label{sec:discussion}

The decomposition in Table~\ref{tab:uncertainty_decomposition_localization} concerns the secondary finite-window magnetic kernel-shape diagnostics and separates empirical form-factor-fit variation from prescribed within-family structural-profile sensitivity. The final localization ensembles use \(N_{\rm rep}=5000\) replicas, matching the data-level propagation. The fit-only ensemble samples the covariance matrix of the selected magnetic form-factor fit while holding the GPD profile fixed. The structural scan instead holds the form-factor fit fixed and varies each continuous profile parameter independently within the baseline \(\pm20\%\) ranges of Table~\ref{tab:profile_variation_ranges}. Its uniform sampling is a numerical coverage prescription rather than a probability distribution inferred from data, so its percentile width is reported only as a structural-sensitivity measure. Table~\ref{tab:uncertainty_decomposition_localization} concerns only the secondary finite-window magnetic kernel-shape diagnostics. With the \(5000\)-replica ensembles, the fit-only percentile half-width exceeds the within-family \(\pm20\%\) structural-scan percentile half-width for all four profile families. The factorized baseline has essentially zero structural contribution because variation of its longitudinal beta parameters does not alter the normalized transverse width. Among the non-factorized families, the Regge-like profile has the largest structural-to-fit ratio, approximately \(0.62\)--\(0.65\), followed by the double-distribution-inspired profile at approximately \(0.47\)--\(0.53\). The correlated exponential profile is substantially less sensitive, at approximately \(0.17\)--\(0.21\). These results refer specifically to within-family variation of the secondary finite-window observables and should not be conflated with the larger discrete spread among the four central profile families or with the slope-defined localization analysis of Section~\ref{sec:localization}.

\begin{table*}[h!]
\centering

\caption{High-statistics decomposition of empirical fit variation and prescribed within-family structural sensitivity for the secondary finite-window magnetic kernel-shape diagnostics. Each entry is the median over \(0.05\le x\le0.95\) of the half-width \((P_{84}-P_{16})/2\), evaluated with \(N_{\rm rep}=5000\). The structural column uses the independent bounded \(\pm20\%\) scan of Table~\ref{tab:profile_variation_ranges} and is not a statistical confidence interval. ``Profile/Fit'' is the ratio of the two reported half-widths. All entries satisfy Profile/Fit \(\le0.80\), so the empirical fit variation exceeds the prescribed within-family structural variation for every row.}
\label{tab:uncertainty_decomposition_localization}
\begin{tabular}{llccc}
\toprule
Profile & Observable & Fit only [fm] & Structural [fm] & Profile/Fit \\
\midrule
Factorized baseline &
\(\sqrt{\mu_2^{\rm win}}\) &
0.015 & 0.000 & 0.00 \\

Factorized baseline &
\(\langle b\rangle^{\rm win}\) &
0.012 & 0.000 & 0.00 \\

Correlated non-factorized &
\(\sqrt{\mu_2^{\rm win}}\) &
0.015 & 0.002 & 0.17 \\

Correlated non-factorized &
\(\langle b\rangle^{\rm win}\) &
0.012 & 0.003 & 0.21 \\

Regge-like &
\(\sqrt{\mu_2^{\rm win}}\) &
0.014 & 0.009 & 0.62 \\

Regge-like &
\(\langle b\rangle^{\rm win}\) &
0.013 & 0.008 & 0.65 \\

Double-distribution-inspired &
\(\sqrt{\mu_2^{\rm win}}\) &
0.014 & 0.007 & 0.47 \\

Double-distribution-inspired &
\(\langle b\rangle^{\rm win}\) &
0.012 & 0.007 & 0.53 \\
\bottomrule
\end{tabular}

\end{table*}

A dedicated replica-count convergence test was performed at \(N_{\rm rep}=80,\ 500,\ 1000,\ 2500,\) and \(5000\). Relative to the final \(5000\)-replica result, the maximum discrepancy decreases from \(11.9\%\) at \(N_{\rm rep}=80\) to \(6.2\%\) at \(500\), \(3.8\%\) at \(1000\), and \(1.4\%\) at \(2500\). The reported uncertainty decomposition therefore uses the \(5000\)-replica ensembles and quotes the widths to \(0.001~\mathrm{fm}\). The dependence on the prescribed structural range was tested independently using \(\pm10\%\), \(\pm20\%\), and \(\pm30\%\) scans. Table~\ref{tab:profile_range_sensitivity} summarizes the median full-envelope half-width of \(b_{\rm rms}^{\rm slope}(x)\) over the evaluated \(x\) interval. The factorized profile is insensitive to these longitudinal-shape variations to numerical precision. For all three transition channels, widening the prescribed range increases the structural envelope smoothly while preserving the qualitative ordering. The correlated exponential profile remains least sensitive, while the DD-inspired profile gives the largest median structural envelope and the Regge-like profile is intermediate for the analytic slope observable. The reconstructed profile dependence is therefore not an artifact of the baseline \(\pm20\%\) choice.

\begin{table*}[h!]
\centering

\caption{Sensitivity of the slope-defined localization scale to the prescribed within-family profile-parameter range. Entries are the median full structural-envelope half-width of \(b_{\rm rms}^{\rm slope}(x)\), in fm, over the evaluated \(x\) interval. Continuous profile parameters are varied independently with uniform coverage of the stated range and no imposed correlations. These values quantify structural sensitivity and are not statistical uncertainties. The factorized profile is zero to numerical precision for all three ranges.}
\label{tab:profile_range_sensitivity}
\begin{tabular}{llccc}
\toprule
Channel & Profile & \(\pm10\%\) & \(\pm20\%\) & \(\pm30\%\) \\
\midrule
\(M\) & Correlated exponential & 0.0033 & 0.0065 & 0.0094 \\
\(M\) & Regge-like             & 0.0082 & 0.0159 & 0.0252 \\
\(M\) & DD-inspired            & 0.0101 & 0.0209 & 0.0324 \\
\midrule
\(E\) & Correlated exponential & 0.0026 & 0.0055 & 0.0078 \\
\(E\) & Regge-like             & 0.0072 & 0.0140 & 0.0172 \\
\(E\) & DD-inspired            & 0.0082 & 0.0173 & 0.0270 \\
\midrule
\(C\) & Correlated exponential & 0.0018 & 0.0036 & 0.0056 \\
\(C\) & Regge-like             & 0.0049 & 0.0090 & 0.0147 \\
\(C\) & DD-inspired            & 0.0060 & 0.0128 & 0.0187 \\
\bottomrule
\end{tabular}

\end{table*}

\subsection{Covariance-prescription sensitivity}

Two covariance issues were tested independently. First, the published model uncertainties were changed from the central point-by-point quadrature prescription to the limiting case where the model component within each input table is fully correlated across \(Q^2\). To isolate this covariance effect from a simultaneous change of functional form, the selected central fit form was retained in each channel when evaluating parameter and localization shifts. Table~\ref{tab:model_error_correlation_sensitivity} summarizes the result. The magnetic modified-dipole parameters change by approximately \(1.8\)--\(2.4\%\), with shifts below \(0.5\) standard deviations of the central fit. For the correlated non-factorized profile, the corresponding maximum changes over \(0.05\le x\le0.95\) are \(1.19\%\) in \(b_{\rm rms}^{\rm slope}(x)\) and \(1.09\%\) in \(\langle b\rangle^{\rm win}(x)\). The electric dipole sector is more sensitive where \(A_{E,0}\) increases by \(20.8\%\) and \(\Lambda_E^2\) decreases by \(19.0\%\) with shifts of approximately one central-fit standard deviation, producing maximum changes of \(11.58\%\) and \(10.51\%\) in the two localization diagnostics. In the Coulomb sector, the fully correlated prescription retains the cubic \(z\) expansion as the AICc-selected form. Holding that functional form fixed, the shifts of \(a_0\), \(a_1\), \(a_2\), and \(a_3\) correspond to approximately \(0.32\), \(0.36\), \(0.38\), and \(0.40\) standard deviations of their central-fit uncertainties, respectively. Relative percentage shifts are not especially informative for \(a_3\) because its central value is close to zero. The corresponding maximum localization changes are \(3.13\%\) in \(b_{\rm rms}^{\rm slope}(x)\) and \(2.70\%\) in \(\langle b\rangle^{\rm win}(x)\).

The second test retains the central point-to-point uncertainty prescription but performs a joint \(M/E/C\) fit using the complete cross-channel covariance matrix obtained from the common replica ensemble. Relative to the independent channel fits, the fitted form factors change by at most \(0.45\%\) for \(G_M^*\), \(1.62\%\) for \(G_E^*\), and \(7.01\%\) for \(G_C^*\) over the representative \(Q^2=0,\ 0.5,\ 1,\ 2,\ 4,\) and \(6~\mathrm{GeV}^2\) points. More directly, the ratios \(G_E^*/G_M^*\) and \(G_C^*/G_M^*\) change by at most \(1.16\%\) and \(7.27\%\), respectively. The largest Coulomb effect occurs at the upper end of the measured momentum-transfer range. This full-block test was also propagated into the multipole-resolved kernels. For the representative \(x\) values used in Section~\ref{sec:multipole_kernels}, the maximum changes in the raw kernel peak magnitude are \(0.05\%\), \(1.30\%\), and \(0.71\%\) for the magnetic, Coulomb, and electric sectors, respectively. After the per-channel shape normalization used in Figure~\ref{fig:multipole_kernels}, the largest pointwise differences are \(0.11\%\), \(1.45\%\), and \(0.50\%\), respectively. Thus the cross-channel covariance has its largest effect in the Coulomb sector but does not alter the qualitative radial-shape comparison shown in Section~\ref{sec:multipole_kernels}.

\begin{table*}[t]
\centering
\caption{Sensitivity to the point-to-point treatment of the published model uncertainties. The central analysis combines statistical and model uncertainties in quadrature point by point. The alternative prescription treats the model component within each input table as fully correlated. Parameter shifts compare the same central functional form under the two prescriptions. For the Coulomb \(z\)-expansion, shifts are quoted in units of the central-fit parameter standard deviation because the highest coefficient is close to zero and a relative percentage shift is not informative. The localization columns give the maximum absolute relative change over \(0.05\le x\le0.95\) for the central correlated non-factorized profile.}
\label{tab:model_error_correlation_sensitivity}
\resizebox{\textwidth}{!}{%
\begin{tabular}{llllcc}
\toprule
Channel & Form & Parameter shifts &
Fully correlated selection &
\(\max|\Delta b_{\rm rms}^{\rm slope}|/b_{\rm rms}^{\rm slope}\) &
\(\max|\Delta\langle b\rangle^{\rm win}|/\langle b\rangle^{\rm win}\) \\
\midrule
\(M\) &
modified dipole &
\(A_0:+1.83\%,\ \Lambda^2:-2.34\%,\ c_1:-2.38\%\) &
modified dipole &
\(1.19\%\) &
\(1.09\%\) \\

\(E\) &
dipole &
\(A_0:+20.83\%,\ \Lambda^2:-19.05\%\) &
dipole &
\(11.58\%\) &
\(10.51\%\) \\

\(C\) &
cubic \(z\) &
\(|\Delta p|/\sigma=0.32,\ 0.36,\ 0.38,\ 0.40\) &
cubic \(z\) &
\(3.13\%\) &
\(2.70\%\) \\
\bottomrule
\end{tabular}
}
\end{table*}

\subsection{Higher transverse-shape moments}

The slope-defined width supplies the primary localization scale, but it does not describe the asymmetry or tail structure of the resolved kernel core. For that purpose the finite-window distribution \(P_{\alpha,p}^{\rm win}(b|x)\) of Eq.~\eqref{eq:windowed_radial_distribution} is used. The windowed mean radius is \(\mu_1^{\rm win}\), while the corresponding variance is
\begin{equation}
\left[\sigma_{\alpha,p}^{\rm win}(x)\right]^2
=
\mu_{2,\alpha,p}^{\rm win}(x)
-
\left[\mu_{1,\alpha,p}^{\rm win}(x)\right]^2.
\label{eq:windowed_variance}
\end{equation}
The finite-window skewness and kurtosis are then
\begin{align}
\gamma_{1,\alpha,p}^{\rm win}(x)
&=
\frac{
\displaystyle
\int_0^{b_{\rm win}}db\,
\left[b-\mu_{1,\alpha,p}^{\rm win}(x)\right]^3
P_{\alpha,p}^{\rm win}(b|x)
}
{
\left[\sigma_{\alpha,p}^{\rm win}(x)\right]^3
},
\\
\kappa_{\alpha,p}^{\rm win}(x)
&=
\frac{
\displaystyle
\int_0^{b_{\rm win}}db\,
\left[b-\mu_{1,\alpha,p}^{\rm win}(x)\right]^4
P_{\alpha,p}^{\rm win}(b|x)
}
{
\left[\sigma_{\alpha,p}^{\rm win}(x)\right]^4
}.
\label{eq:skew_kurt}
\end{align}
The reported kurtosis is the ordinary fourth standardized moment rather than the excess kurtosis. Because these quantities are evaluated over the fixed \(b_{\rm win}=2.5~\mathrm{fm}\) analysis window, they are finite-window core-shape diagnostics rather than moments of the band-limited transform extended to infinite radius. They are therefore used to compare the shapes generated by different profile families, not to define the primary transverse size.

\begin{figure*}[h!]
    \centering
    \includegraphics[width=\textwidth]{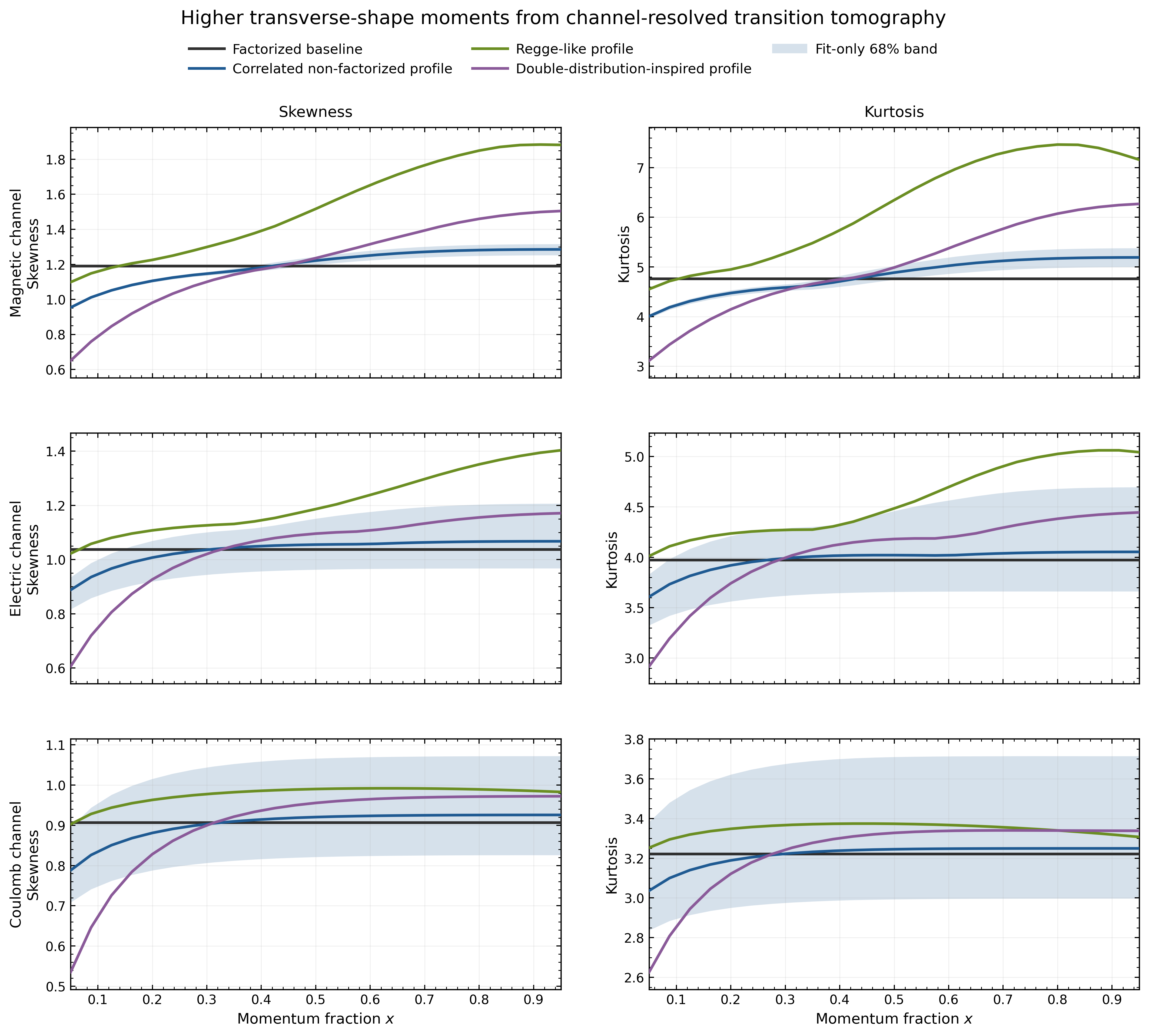}
    \caption{Finite-window higher transverse-shape diagnostics for the magnetic, electric, and Coulomb channels. The left column shows \(\gamma_1^{\rm win}(x)\) and the right column shows \(\kappa^{\rm win}(x)\), both evaluated from the magnitude-weighted radial distribution over \(0\le b\le2.5~\mathrm{fm}\). The four curves show the central profile families. The blue shaded region gives the 16th--84th percentile fit-only uncertainty for the central correlated exponential profile, obtained from \(5000\) replicas of the covariance matrix of the selected transition-form-factor fit in each channel with the GPD profile held fixed. Structural-profile variation is not included in the shaded band and is represented separately by the spread among the central profile-family curves. These quantities characterize the resolved core shape of the band-limited transition kernel and are not interpreted as infinite-range moments.}
    \label{fig:skewness_kurtosis}
\end{figure*}

Figure~\ref{fig:skewness_kurtosis} shows that the finite-window higher-shape diagnostics retain substantial sensitivity to the assumed \(x\)--\(t\) profile even when the primary slope-defined localization scales are similar. The factorized baseline is \(x\)-independent in all three channels, whereas each non-factorized family produces a characteristic \(x\) dependence. In the magnetic channel, both skewness and kurtosis increase strongly with \(x\) for the Regge-like and DD-inspired profiles, with the Regge-like profile giving the largest values over most of the intermediate- and large-\(x\) region. The correlated exponential profile varies more moderately and approaches a comparatively smooth high-\(x\) plateau. The electric channel shows the same broad central-profile ordering with a smaller separation, while the Coulomb central curves become more tightly clustered at moderate and large \(x\). For the Coulomb Regge-like profile the higher moments rise and then flatten or decrease mildly toward the largest \(x\) values. Thus the largest separation among the central higher-moment curves occurs in the magnetic channel. The fit-only uncertainty bands add an important qualification to this structural comparison. In the magnetic channel the selected-form-factor uncertainty remains comparatively narrow, so the large intermediate- and high-\(x\) separation of the Regge-like and DD-inspired curves from the correlated exponential result is predominantly structural. In the electric channel the fit uncertainty is broader and overlaps a substantial fraction of the central profile-family spread. In the Coulomb channel the fit-only band encompasses most of the central-profile variation over much of the displayed \(x\) range, showing that the higher-shape diagnostics there are limited more strongly by the empirical form-factor uncertainty. Structural-profile variation is not included in these bands, so empirical fit uncertainty and the discrete dependence on profile family remain separately identifiable. Because the moments are constructed from the magnitude-weighted kernel over the fixed interval \(0\le b\le2.5~\mathrm{fm}\), they are finite-window diagnostics of the resolved kernel core rather than global moments or direct empirical measurements of transverse-density asymmetry.

\subsection{Physical interpretation}

The interpretation of the reconstructed transverse structures should be made in light of the recent formal transition-GPD literature. The complete \(N\to\Delta\) transition-GPD basis contains more independent structures than are fixed by the multipole form-factor input alone~\cite{Kim:2025CompleteDefinition}. In addition, impact-parameter transition densities are non-diagonal matrix elements between distinct hadron states and require a specified light-front spin basis for a complete physical density interpretation~\cite{Kim:2023QCDAM,Kim:2025TensorPolarized,Kim:2026MultipoleStructure}. The results below are therefore interpreted as form-factor-constrained radial diagnostics of modeled transition-GPD profiles, not as a complete extraction of the full \(N\to\Delta\) light-front transition density.

For the slope-defined localization observable, the appearance of \(x\) dependence is a direct consequence of the chosen non-factorized profile ansatz. Equation~\eqref{eq:slope_closed_form} separates this modeled contribution from the empirical transition-form-factor slope where \(G_\alpha^{*\prime}(0)/G_\alpha^*(0)\) fixes an \(x\)-independent channel offset, while \(g_p(x)-\langle g_p\rangle_p\) fixes the profile-dependent \(x\) variation. The factorized profile therefore gives an \(x\)-independent width by construction. The non-factorized profiles provide controlled alternative \(x\)-dependent spatial hypotheses under the same form-factor normalization rather than a data-driven extraction of that \(x\) dependence.

The magnetic sector provides the cleanest illustration of this separation. It is the dominant transition channel, its form factor is most stable under model selection, and its tomography maps show the most controlled profile dependence. The correlated non-factorized, Regge-like, and double-distribution-inspired profiles all use the same fitted \(G_M^*(Q^2)\) and hence the same imposed normalization \(F_M(t)=2G_M^*(-t)\), but they produce different \(x\)-dependent slope-defined widths and different finite-window core-shape behavior. Therefore the fitted \(G_M^*(Q^2)\), through \(F_M(t)\), fixes the integrated transition strength but does not fix the \(x\)-resolved spatial structure of the modeled transition kernel. The transition tomography is consequently best understood as a form-factor-constrained spatial reconstruction with explicit structural-profile uncertainty. The electric and Coulomb sectors extend this interpretation to the deformation-sensitive components of the \(N\to\Delta(1232)\) transition. Their absolute magnitudes are smaller and their form-factor descriptions are more sensitive to the ratio-sector inputs and model selection. The electric and Coulomb channels both retain stronger model dependence than the dominant magnetic normalization would suggest from transition strength alone, but the data-domain-limited reconstruction removes the artificial abrupt Coulomb behavior generated by the high-\(Q^2\) extrapolation. This hierarchy is physically reasonable. The magnetic transition is the dominant and most directly constrained component, while the electric and Coulomb amplitudes probe smaller deformation and longitudinal-response effects that are more sensitive to modeling choices and low-\(Q^2\) behavior~\cite{Pascalutsa:2006up,Ramalho:2016buz}.

The large-\(N_c\) comparison provides an independent theoretical reference for this structural freedom. Among the present profile set, the Regge-like form lies closest to the normalized \(E_v^u-E_v^d\) magnetic benchmark over the tested momentum-transfer range. This comparison is not used as an additional fit criterion or to promote the Regge-like profile to a unique physical solution. Rather, it provides an external theoretical reference for assessing the otherwise underconstrained longitudinal \(x\)-dependence.

The multipole-resolved radial kernels in Section~\ref{sec:multipole_kernels} reinforce the same conclusion from a different angle. The \(M\), \(C\), and \(E\) sectors do not simply reproduce the same transverse shape with different normalization. In the displayed shape-normalized comparison, the magnetic \(J_0\) kernel is positive and concentrated toward the smallest \(b\), followed by a monotonic decrease. The Coulomb \(J_1\) kernel is suppressed near the origin and reaches a positive maximum at finite \(b\). The electric \(J_2\) kernel is more strongly suppressed near the origin and reaches a broader maximum at larger transverse radius. Within the fixed band-limited production transform at \(\Delta_{\max}=\sqrt6~\mathrm{GeV}\), the characteristic \(C\) and \(E\) peak positions move inward as \(x\) increases, and none of the three displayed normalized kernels changes sign over the plotted interval. The kernel representation is deliberately convention-safe as it avoids claiming a full polarized transverse charge or current density, while still displaying how the deformation-sensitive multipoles populate transverse radius. This is the appropriate level of interpretation for the present reconstruction. Relative to the precursor study~\cite{Marinaro:2025TransitionGPD}, the present work makes two substantive advances. First, it uses model selection to define the \(Q^2\)-dependent form-factor layer before introducing the GPD profiles. Second, it replaces the purely factorized transition-GPD ansatz with normalized non-factorized profile families whose common form-factor normalization is imposed analytically, allowing the consequences of different \(x\)--\(t\) correlations to be compared without changing the integrated transition strength. These advances are complementary to, rather than substitutes for, the complete formal transition-GPD definitions and multipole decompositions developed in Refs.~\cite{Kim:2025CompleteDefinition,Kim:2026MultipoleStructure}. The present analysis supplies a data-driven and reproducible phenomenological pipeline. The formal literature supplies the operator-level basis and the fully specified light-front density interpretation. All three non-factorized profile families impose smooth \(x\)-dependent slope localization, but the detailed form of that dependence is a property of the respective ansatz rather than an independently data-constrained result. Because the profile contribution to \([b_{\rm rms}^{\rm slope}]^2\) is common among \(M\), \(E\), and \(C\), the electric and Coulomb channels do not provide independent tests of this particular \(x\)-dependent slope structure. Their complementary deformation sensitivity instead appears in the explicit multipole-resolved kernels and finite-window radial-shape diagnostics. The magnetic channel provides the most stable spatial interpretation, while the electric and Coulomb sectors supply deformation-sensitive diagnostics whose profile and covariance-prescription dependence should be retained rather than hidden. In particular, the fully correlated model-error cross-check shows a non-negligible sensitivity of the electric localization scale, whereas the full cross-channel covariance produces only modest changes in the relative multipole normalization and leaves the shape-normalized kernel comparison qualitatively unchanged. This balance between robust qualitative conclusions and explicit structural uncertainty is a central result of the analysis.

\section{Conclusions}
\label{sec:summary}

This work develops a form-factor-constrained, non-factorized transition-GPD reconstruction for the \(N\to\Delta(1232)\) electromagnetic transition. The published CLAS magnetic multipole amplitude and quadrupole ratios are propagated replica by replica, with the derived electric and scalar/Coulomb multipoles converted to the dimensionless Jones--Scadron form factors \(G_M^*\), \(G_E^*\), and \(G_C^*\). AICc selects a modified dipole, dipole, and cubic \(z\) expansion for the magnetic, electric, and Coulomb sectors, respectively. These fits determine the empirical momentum-transfer dependence, while the longitudinal structure is represented by factorized, correlated exponential, Regge-like, and DD-inspired profile families under the common imposed positive-\(x\) normalization \(F_\alpha(t)=2G_\alpha^*(-t)\). The upper boundary of the production impact-parameter transform is fixed by the largest measured absolute-form-factor point, \(Q^2=6~\mathrm{GeV}^2\), so that no high-\(Q^2\) continuation beyond the measured range enters the production reconstruction. The cutoff study demonstrates that extending the empirical fits far beyond this domain can produce substantial Coulomb-sector sensitivity, whereas the fixed-domain production quadrature is numerically converged.

The analytic slope relation shows that the factorized construction gives an \(x\)-independent width by definition, while each non-factorized ansatz imposes a smooth and distinct modeled \(x\) dependence. The empirical transition form factors determine channel-dependent but \(x\)-independent offsets in the squared width. The shape-normalized multipole kernels resolve correspondingly different radial structures. The magnetic \(J_0\) kernel is concentrated toward the smallest displayed impact parameter and decreases monotonically, the Coulomb \(J_1\) kernel has a positive finite-radius maximum, and the electric \(J_2\) kernel has a broader maximum at still larger transverse radius. Within the band-limited production transform at \(\Delta_{\max}=\sqrt6~\mathrm{GeV}\), the characteristic Coulomb and electric peak positions move inward as \(x\) increases, and none of the displayed normalized kernels changes sign over the plotted interval. These smooth resolved-core structures should be distinguished from the weak large-\(b\) ringing generated by the sharp transverse-momentum cutoff. That cutoff-induced ringing prevents magnitude-weighted radial moments from being assigned stable infinite-range values and is why the small-\(Q^2\) slope is retained as the primary localization observable. Mean radius, skewness, and kurtosis are therefore used only as finite-window diagnostics of the resolved kernel core. Figure~\ref{fig:skewness_kurtosis} further shows that the higher-shape moments retain substantial profile-family dependence, with the largest central-profile separation occurring in the magnetic channel over much of the displayed \(x\) range. The comparatively narrow magnetic fit-only band leaves much of this separation clearly structural, whereas the broader electric and especially Coulomb bands show that empirical form-factor uncertainty becomes comparable to, or larger than, much of the corresponding central profile-family spread. The magnetic form factor remains the most tightly constrained empirical normalization, while covariance-prescription sensitivity is largest in the electric localization sector, where the fully correlated model-error test changes the localization scale by approximately \(10\)--\(12\%\). The joint cross-channel covariance test leaves the shape-normalized multipole kernels stable at approximately the \(1.5\%\) level or better. The large-\(N_c\) comparison provides an independent reference for the unresolved longitudinal structure, with the Regge-like ansatz qualitatively closest among the four tested profiles but not uniquely selected. The principal limitations are that the longitudinal \(x\)-dependence is modeled rather than extracted directly and that the \(Q^2=0\) form-factor slope inherits the selected empirical parametrization below the lowest absolute input point at \(Q^2=0.3~\mathrm{GeV}^2\). Future measurements and global analyses that constrain this dependence within the complete \(N\to\Delta\) transition-GPD operator basis will be required for a more model-independent spatial reconstruction.

\section*{Acknowledgments}

This work gratefully acknowledges the CLAS Collaboration and Hall B researchers for the published \(\Delta(1232)\) electroproduction results that form the empirical basis of this analysis. Acknowledgment also goes to the broader Jefferson Lab scientific mission and community, whose experimental and phenomenological work has made detailed studies of the \(N\to\Delta(1232)\) transition possible. This work used publicly available data.

%\section*{Declarations}

%\subsection*{Funding and competing interests}
%This research received no support from any funding agency. There are no competing interests.

%\section*{Data availability statement}
%The empirical electroproduction and quadrupole-ratio inputs used in this work are publicly available from the sources cited in the manuscript. The processed numerical tables, figure data, diagnostic outputs, and other numerical products supporting the results of this study are available from the corresponding author upon request. The analysis source code is not publicly deposited.

%\subsection*{Author contributions}
%The author(s) conceived the study, implemented the numerical analysis, produced the figures and tables, interpreted the results, and wrote the manuscript.

%\subsection*{Ethics approval and consent}
%Not applicable.

\end{document}